\documentclass[aps,prd,reprint,preprintnumbers,showpacs,floatfix,nofootinbib,superscript address]{revtex4-2}
\usepackage[utf8]{inputenc}
\usepackage{parskip}
\usepackage{amssymb}
\usepackage{stix}
\usepackage{hhline}
\usepackage{amsmath}
\usepackage{mathtools}
\usepackage[dvipsnames]{xcolor}
\usepackage{xspace}
\usepackage{multirow,tabularx}
\usepackage{siunitx}
\usepackage{multirow}
\usepackage{graphicx}
\usepackage{xstring}
\usepackage{etoolbox}
\usepackage{notoccite}
\usepackage{tocbasic}
\usepackage{natbib}
\usepackage{mathrsfs}
\usepackage{lineno}
\usepackage{tensor}
\usepackage{accents}
\usepackage{tikz}
\usepackage{listings}
\allowdisplaybreaks

\usepackage{scalerel,stackengine}
\usepackage{upgreek}

\usepackage{enumitem}

\newrobustcmd{\TimeFunc}{	\mathscr{T}\xspace
}
\newrobustcmd{\AxiTimeFunc}{	\mathscr{S}\xspace
}
\newrobustcmd{\SpaceFunc}{	\mathscr{R}\xspace
}
\newrobustcmd{\MixTPhi}{	\mathscr{M}_\phi\xspace
}
\newrobustcmd{\MixTheta}{	\mathscr{M}_\theta\xspace
}
\newrobustcmd{\InduceH}{	\mathscr{H}\xspace
}

\newrobustcmd{\Planck}{%
	{M_{\text{Pl}}}%
}

\newrobustcmd{\rD}[1]{%
	\tensor{\mathring{\nabla}}{#1}%
}

\newrobustcmd{\rcD}[1]{%
	\tensor{\nabla}{#1}%
}

\newrobustcmd{\rR}[1]{%
	\tensor{\mathring{R}}{#1}%
}

\newrobustcmd{\rcR}[1]{%
	\tensor{R}{#1}%
}

\newrobustcmd{\T}[2][placeholder]{%
	\IfEqCase{#1}{%
	{placeholder}{\tensor{T}{#2}}%
	{1}{\tensor[^{(1)}]{T}{#2}}%
	{2}{\tensor[^{(2)}]{T}{#2}}%
	{3}{\tensor[^{(3)}]{T}{#2}}%
	}%
	[\packageError{cosmicclass}{Symbol #1 is not an irreducible part!}{}]%
}

\newrobustcmd{\TLambda}[2][placeholder]{%
	\IfEqCase{#1}{%
	{placeholder}{\tensor{\lambda}{#2}}%
	{1}{\tensor[^{(1)}]{\lambda}{#2}}%
	{2}{\tensor[^{(2)}]{\lambda}{#2}}%
	{3}{\tensor[^{(3)}]{\lambda}{#2}}%
	}%
	[\packageError{cosmicclass}{Symbol #1 is not an irreducible part!}{}]%
}

\newrobustcmd{\States}[4]{%
	{\tensor*{#4}{^#3_{#1^#2}}}%
}
\newrobustcmd{\FieldIndices}[2]{%
	{#2_#1}%
}
\newrobustcmd{\ParallelFieldIndices}[3]{%
	{{\overline{#3}}_{#1^#2}}%
}

\newrobustcmd{\FieldUp}[2]{%
	{\tensor{\zeta}{^{\FieldIndices{#1}{#2}}}}%
}
\newrobustcmd{\FieldDown}[2]{%
	{\tensor{\zeta}{_{\FieldIndices{#1}{#2}}}}%
}
\newrobustcmd{\SourceUp}[2]{%
	{\tensor{j}{^{\FieldIndices{#1}{#2}}}}%
}
\newrobustcmd{\SourceDown}[2]{%
	{\tensor{j}{_{\FieldIndices{#1}{#2}}}}%
}

\newrobustcmd{\FieldUpState}[5]{%
	{\tensor{\zeta\big(\States{#1}{#2}{#3}{#4}\big)}{^{\ParallelFieldIndices{#1}{#2}{#5}}}}%
}
\newrobustcmd{\SourceUpFullState}[5]{%
	{\tensor{j\big(\States{#1}{#2}{#3}{#4}\big)}{^{\FieldIndices{#3}{#5}}}}%
}
\newrobustcmd{\FieldDownState}[5]{%
	{\tensor{\zeta\big(\States{#1}{#2}{#3}{#4}\big)}{_{\ParallelFieldIndices{#1}{#2}{#5}}}}%
}
\newrobustcmd{\FieldDownFullState}[5]{%
	{\tensor{\zeta\big(\States{#1}{#2}{#3}{#4}\big)}{_{\FieldIndices{#3}{#5}}}}%
}

\newrobustcmd{\WaveOperatorTensorUpDown}[4]{%
	{\tensor{\mathcal{O}}{^{\FieldIndices{#1}{#3}}_{\FieldIndices{#2}{#4}}}}%
}

\newrobustcmd{\Normalisation}[4]{%
	{c\big(\States{#1}{#2}{#3}{#4}\big)}%
}
\newrobustcmd{\SPODownUp}[8]{%
	{\tensor{%
	\mathcal{P}%
	\big(\States{#1}{#2}{#3}{#5},\States{#1}{#2}{#4}{#6}\big)%
	}{_{\FieldIndices{#3}{#7}}^{\FieldIndices{#4}{#8}}}}%
}
\newrobustcmd{\SPOUpDown}[8]{%
	{\tensor{%
	\mathcal{P}%
	\big(\States{#1}{#2}{#3}{#5},\States{#1}{#2}{#4}{#6}\big)%
	}{^{\FieldIndices{#3}{#7}}_{\FieldIndices{#4}{#8}}}}%
}
\newrobustcmd{\ReducedSPODownUp}[6]{%
	{\tensor{%
	\mathcal{P}%
	\big(\States{#1}{#2}{#3}{#4}\big)%
	}{_{\ParallelFieldIndices{#1}{#2}{#5}}^{\FieldIndices{#3}{#6}}}}%
}
\newrobustcmd{\ReducedSPOUpDown}[6]{%
	{\tensor{%
	\mathcal{P}%
	\big(\States{#1}{#2}{#3}{#4}\big)%
	}{^{\ParallelFieldIndices{#1}{#2}{#5}}_{\FieldIndices{#3}{#6}}}}%
}

\newrobustcmd{\GaugeVarying}[3]{%
	{\tensor*{g}{_{#3_{#1^#2}}}}%
}
\newrobustcmd{\GaugeVaryingConj}[3]{%
	{\tensor*{g}{^*_{#3_{#1^#2}}}}%
}
\newrobustcmd{\GaugeFixing}[3]{%
	{\tensor*{h}{_{#3_{#1^#2}}}}%
}
\newrobustcmd{\Mass}[3]{%
	{\tensor*{m}{_{#3_{#1^#2}}}}%
}
\newrobustcmd{\SquareMass}[3]{%
	{\tensor*{m}{^2_{#3_{#1^#2}}}}%
}
\newrobustcmd{\NullVectors}[3]{%
	{#3_{#1^#2}}%
}
\newrobustcmd{\NullVector}[3]{%
	{\tensor*{\mathsf{v}}{_{#3_{#1^#2}}}}%
}

\newrobustcmd{\NullVectorComponents}[5]{%
	{\tensor{\left[\NullVector{#1}{#2}{#3}\right]}{_{\States{#1}{#2}{#4}{#5}}}}%
}

\newrobustcmd{\SpinMin}[0]{%
	{J_{\mathrm{F}}}%
}
\newrobustcmd{\SpinMax}[0]{%
	{J_{\mathrm{L}}}%
}

\newrobustcmd{\ParityMin}[0]{%
	{P_{\mathrm{F}}}%
}
\newrobustcmd{\ParityMax}[0]{%
	{P_{\mathrm{L}}}%
}

\newrobustcmd{\FieldMin}[0]{%
	{X_{\mathrm{F}}}%
}
\newrobustcmd{\FieldMax}[0]{%
	{X_{\mathrm{L}}}%
}

\newrobustcmd{\VectorMin}[0]{%
	{a_{\mathrm{F}}}%
}
\newrobustcmd{\VectorMax}[0]{%
	{a_{\mathrm{L}}}%
}

\newrobustcmd{\ZeroVector}[0]{%
	{\mathsf{0}}%
}
\newrobustcmd{\ConstraintMatrix}[0]{%
	{\mathsf{C}}%
}
\newrobustcmd{\SourceVector}[0]{%
	{\mathsf{J}}%
}

\newrobustcmd{\NullVectorConj}[3]{%
	{\tensor*{\mathsf{v}}{^\dagger_{#3_{#1^#2}}}}%
}

\newrobustcmd{\WaveOperator}[0]{%
	{\tensor{\mathsf{O}}{}}%
}
\newrobustcmd{\WaveOperatorConj}[0]{%
	{\tensor{\mathsf{O}}{^\dagger}}%
}
\newrobustcmd{\MoorePenrose}[0]{%
	{\tensor{\mathsf{O}}{^+}}%
}
\newrobustcmd{\MoorePenroseJP}[2]{%
	{\tensor*{\mathsf{O}}{^+_{#1^#2}}}%
}
\newrobustcmd{\WaveOperatorJP}[2]{%
	{\tensor*{\mathsf{O}}{_{#1^#2}}}%
}
\newrobustcmd{\WaveOperatorJPComponents}[6]{%
	{\tensor{\left[\WaveOperatorJP{#1}{#2}\right]}{_{\States{#1}{#2}{#3}{#5}\States{#1}{#2}{#4}{#6}}}}%
}
\newrobustcmd{\Propagator}[0]{%
	{\tensor{\mathsf{P}}{}}%
}
\newrobustcmd{\Similarity}[0]{%
	{\tensor{\mathsf{U}}{}}%
}

\newrobustcmd{\Source}[0]{%
	{\tensor{\mathsf{j}}{}}%
}
\newrobustcmd{\SourceConj}[0]{%
	{\tensor{\mathsf{j}}{^\dagger}}%
}
\newrobustcmd{\Field}[0]{%
	{\tensor{\upzeta}{}}%
}
\newrobustcmd{\FieldConj}[0]{%
	{\tensor{\upzeta}{^\dagger}}%
}

\newrobustcmd{\Powers}[3]{%
	{#3_{#1^#2}}%
}
\newrobustcmd{\Masses}[3]{%
	{#3_{#1^#2}}%
}

\newrobustcmd{\NewRes}[2]{%
	{\mathop{\mathrm{Res}}_{#1\mapsto#2}}%
}

\newrobustcmd{\DerOp}[2]{%
	\tensor[^{(#1)}]{\bar{\mathscr{D}}}{#2}%
}
\newrobustcmd{\bB}[1]{%
	\tensor{\bar{B}}{#1}%
}
\newrobustcmd{\bG}[1]{%
	\tensor*{\bar{G}}{#1}%
}
\newrobustcmd{\bg}[1]{%
	\tensor*{\bar{g}}{#1}%
}
\newrobustcmd{\bR}[1]{%
	\tensor*{\bar{R}}{#1}%
}
\newrobustcmd{\bD}[1]{%
	\tensor{\overline{\nabla}}{#1}%
}
\newrobustcmd{\hp}[1]{%
	\tensor*{h}{#1}%
}
\newrobustcmd{\hpb}[1]{%
	\tensor*{\bar{h}}{#1}%
}

\newrobustcmd{\Sources}[1]{%
	\hat{j}%
}
\newrobustcmd{\FieldsT}[1]{%
	\hat{\zeta}^{\text{T}}%
}

\newrobustcmd{\AMat}[1]{%
	{\mathsf{A}}%
}
\newrobustcmd{\AMatT}[1]{%
	{\mathsf{A}}^{\text{T}}%
}
\newrobustcmd{\AMatU}[1]{%
	\tensor[^{\text{s}}]{\mathsf{A}}{}%
}
\newrobustcmd{\AMatUt}[1]{%
	\tensor[^{\text{s}}]{\tilde{\mathsf{A}}}{}%
}
\newrobustcmd{\AMatO}[1]{%
	\tensor[^{\text{a}}]{\mathsf{A}}{}%
}

\newrobustcmd{\BMat}[1]{%
	{\mathsf{B}}%
}
\newrobustcmd{\BMatT}[1]{%
	{\mathsf{B}}^{\text{T}}%
}
\newrobustcmd{\BMatU}[1]{%
	\tensor[^{\text{s}}]{\mathsf{B}}{}%
}
\newrobustcmd{\BMatO}[1]{%
	\tensor[^{\text{a}}]{\mathsf{B}}{}%
}
\newrobustcmd{\BMatOt}[1]{%
	\tensor[^{\text{a}}]{\tilde{\mathsf{B}}}{}%
}

\newrobustcmd{\CMat}[1]{%
	{\mathsf{C}}%
}
\newrobustcmd{\CMatT}[1]{%
	{\mathsf{C}}^{\text{T}}%
}
\newrobustcmd{\CMatU}[1]{%
	\tensor[^{\text{s}}]{\mathsf{C}}{}%
}
\newrobustcmd{\CMatUt}[1]{%
	\tensor[^{\text{s}}]{\tilde{\mathsf{C}}}{}%
}

\newrobustcmd{\rotate}[1]{%
	\mathsf{R}%
}
\newrobustcmd{\rotateT}[1]{%
	\mathsf{R}^{\text{T}}%
}

\newrobustcmd{\omeg}[1]{%
	\mathsf{S}%
}
\newrobustcmd{\omegT}[1]{%
	\mathsf{S}^{\text{T}}%
}

\newrobustcmd{\uni}[1]{%
	\mathsf{n}_{#1}%
}
\newrobustcmd{\uniT}[1]{%
	\mathsf{n}_{#1}^{\text{T}}%
}
\newrobustcmd{\eig}[1]{%
	\mathsf{v}_{#1}%
}
\newrobustcmd{\eigT}[1]{%
	\mathsf{v}_{#1}^{\text{T}}%
}
\newrobustcmd{\lam}[1]{%
	\lambda_{#1}%
}

\newcommand{\mhDel}[1]{{\color[RGB]{0,100,0}\ifmmode\text{\sout{$#1$}}\else\sout{#1}\fi}}

\newrobustcmd{\pea}[1]{%
	\emph{#1}\textbf{.\ \ \ ---}
}

\newrobustcmd{\typeone}[1]{%
	{I}	
}
\newrobustcmd{\typetwo}[1]{%
	{II}	
}
\newrobustcmd{\typethree}[1]{%
	{III}	
}
\newrobustcmd{\typefour}[1]{%
	{IV}	
}
\newrobustcmd{\typefive}[1]{%
	{V}	
}
\newrobustcmd{\typesix}[1]{%
	{VI}	
}

\newrobustcmd{\first}[1]{%
	{1\textsuperscript{st}}	
}
\newrobustcmd{\second}[1]{%
	{2\textsuperscript{nd}}	
}
\newrobustcmd{\third}[1]{%
	{3\textsuperscript{rd}}	
}
\newrobustcmd{\fourth}[1]{%
	{4\textsuperscript{th}}	
}

\newrobustcmd{\IsOff}[1]{%
	{\ =0\ \land\ }	
}

\newrobustcmd{\MAGg}[1]{%
	\tensor{g}{#1}
}

\newrobustcmd{\MAGd}[1]{%
	\tensor*{\delta}{#1}
}
\newrobustcmd{\MAGl}[1]{%
	\tensor{\xi}{#1}
}

\newrobustcmd{\gflat}[1]{%
  \tensor{\eta}{#1}
}
\newrobustcmd{\h}[1]{%
  \tensor{h}{#1}
}
\newrobustcmd{\MAGA}[1]{%
  \tensor{A}{#1}
}
\newrobustcmd{\MAGF}[1]{%
	\tensor{\mathcal{F}}{#1}
}
\newrobustcmd{\MAGFP}[1]{%
	\tensor{\accentset{P}{\mathcal{F}}}{#1}
}
\newrobustcmd{\MAGFTri}[1]{%
	\tensor{\accentset{\Delta}{\mathcal{F}}}{#1}
}
\newrobustcmd{\MAGFa}[1]{%
	\tensor{\mathcal{F}}{^{(14)}#1}
}
\newrobustcmd{\MAGFb}[1]{%
	\tensor{\mathcal{F}}{^{(13)}#1}
}
\newrobustcmd{\MAGT}[1]{%
	\tensor{\mathcal{T}}{#1}
}
\newrobustcmd{\MAGQ}[1]{%
	\tensor{\mathcal{Q}}{#1}
}
\newrobustcmd{\MAGQt}[1]{%
	\tensor{\tilde{\mathcal{Q}}}{#1}
}

\newrobustcmd{\alp}[1]{%
       {\alpha}	
}

\newrobustcmd{\bet}[1]{%
  \tensor[^{(#1)}]{\mu}{}
}

\stackMath 
\newcommand\OmitIndices[1]{%
\savestack{\tmpbox}{\stretchto{%
\scaleto{%
\scalerel*[\widthof{\ensuremath{#1}}]{\kern-.6pt\curlywedge\kern-.6pt}%
{\rule[-\textheight/2]{1ex}{\textheight}}
}{\textheight}%
}{0.5ex}}%
\stackon[1pt]{#1}{\tmpbox}%
}

\newrobustcmd{\LPV}[1]{%
	\IfEqCase{#1}{%
	{}{L_{\text{PV}}}%
	}%
	[{L_{\text{PV}}\left(#1\right)}]%
}

\newrobustcmd{\F}[2][placeholder]{%
	\IfEqCase{#1}{%
	{placeholder}{\tensor{T}{#2}}%
	{1}{\tensor[^{(1)}]{F}{#2}}%
	{2}{\tensor[^{(2)}]{F}{#2}}%
	{3}{\tensor[^{(3)}]{F}{#2}}%
	}%
	[\packageError{cosmicclass}{Symbol #1 is not an irreducible part!}{}]%
}

\newrobustcmd{\GenericVector}[1]{%
	\smash{{#1}_{\tensor[^{{(2)}}]{\hspace{-1pt}\lambda}{}}^{J^P}}%
}
\newrobustcmd{\GenericTensor}[1]{%
	\smash{{#1}_{\tensor[^{{(1)}}]{\hspace{-1pt}\lambda}{}}^{J^P}}%
}

\newrobustcmd{\g}[1]{%
	\tensor{g}{#1}%
}
\newrobustcmd{\rcCon}[1]{%
	\tensor*{\Gamma}{#1}%
}
\newrobustcmd{\rCon}[1]{%
	\tensor*{\mathring{\Gamma}}{#1}%
}
\newrobustcmd{\B}[1]{%
	\tensor{B}{#1}%
}
\newrobustcmd{\PD}[1]{%
	\tensor{\partial}{#1}%
}
\newrobustcmd{\BConj}[1]{%
	\tensor{B^\dagger}{#1}%
}
\newrobustcmd{\N}[1]{%
	\tensor{n}{#1}%
}
\newrobustcmd{\J}[1]{%
	\tensor{J}{#1}%
}
\newrobustcmd{\En}[1]{%
	\mathcal{E}%
}
\newrobustcmd{\Mo}[1]{%
	p%
}
\newrobustcmd{\JConj}[1]{%
	\tensor{J^\dagger}{#1}%
}
\newrobustcmd{\K}[1]{%
	\tensor{X}{_#1}%
}
\newrobustcmd{\KConj}[1]{%
	\tensor*{X}{^{\dagger}_{#1}}%
}

\renewrobustcmd{\F}[2][placeholder]{%
	\IfEqCase{#1}{%
	{placeholder}{\tensor{T}{#2}}%
	{1}{\tensor[^{(1)}]{F}{#2}}%
	{2}{\tensor[^{(2)}]{F}{#2}}%
	{3}{\tensor[^{(3)}]{F}{#2}}%
	}%
	[\packageError{cosmicclass}{Symbol #1 is not an irreducible part!}{}]%
}

\newrobustcmd{\Si}[2][placeholder]{%
	\IfEqCase{#1}{%
	{placeholder}{\tensor{T}{#2}}%
	{1}{\tensor[^{(1)}]{S}{#2}}%
	{2}{\tensor[^{(2)}]{S}{#2}}%
	{3}{\tensor[^{(3)}]{S}{#2}}%
	}%
	[\packageError{cosmicclass}{Symbol #1 is not an irreducible part!}{}]%
}

\newrobustcmd{\mass}[2][placeholder]{%
	\IfEqCase{#1}{%
	{placeholder}{\tensor{m}{#2}}%
	{1}{\tensor[^{(1)}]{m}{#2}}%
	{2}{\tensor[^{(2)}]{m}{#2}}%
	{3}{\tensor[^{(3)}]{m}{#2}}%
	}%
	[\packageError{cosmicclass}{Symbol #1 is not an irreducible part!}{}]%
}

\newrobustcmd{\RLambda}[2][placeholder]{%
	\IfEqCase{#1}{%
	{placeholder}{\tensor{\lambda}{#2}}%
	{1}{\tensor[^1]{\lambda}{#2}}%
	{2}{\tensor[^2]{\lambda}{#2}}%
	{3}{\tensor[^3]{\lambda}{#2}}%
	{R}{\tensor[^{(R)}]{\lambda}{#2}}%
	}%
	[\packageError{cosmicclass}{Symbol #1 is not an irreducible part!}{}]%
}

\newrobustcmd{\QLambda}[2][placeholder]{%
	\IfEqCase{#1}{%
		{placeholder}{\tensor{\hat{\lambda}}{#2}}%
	{1}{\tensor[^1]{\hat{\lambda}}{#2}}%
	{2}{\tensor[^2]{\hat{\lambda}}{#2}}%
	{3}{\tensor[^3]{\hat{\lambda}}{#2}}%
	}%
	[\packageError{cosmicclass}{Symbol #1 is not an irreducible part!}{}]%
}

\newrobustcmd{\Mgra}[1]{%
  {\tensor{M}{_{\text{#1}}}}%
}
\newrobustcmd{\Mpro}[1]{%
  {\tensor{\mathproper{M}}{_{\text{#1}}}}%
}
\newrobustcmd{\Malt}[1]{%
  {\tensor{\mathscr{M}}{_{\text{#1}}}}%
}
\newrobustcmd{\Mkom}[1]{%
  {\tensor{\mathfrak{M}}{_{\text{#1}}}}%
}

\newrobustcmd{\Mtotal}{%
  {\tensor{M}{_{\text{T}}}}%
}

\newrobustcmd{\Qtotal}{%
  {\tensor{Q}{_{\text{T}}}}%
}

\newrobustcmd{\Qtotalcal}{%
  {\tensor{\mathcal{  Q}}{_{\text{T}}}}%
}

\newrobustcmd{\action}[1]{%
  {\tensor{S}{_{\text{#1}}}}%
}

\newrobustcmd{\lagrangian}[1]{%
  {\tensor{L}{_{\text{#1}}}}%
}

\newrobustcmd{\lagrangianprop}[1]{%
  {\tensor{\mathproper{L}}{_{\text{#1}}}}%
}

\newrobustcmd{\epl}{%
  {\tensor{\mathsf{e}}{_+}}%
}

\newrobustcmd{\epe}{%
  {\tensor{\mathsf{e}}{_\perp}}%
}

\newrobustcmd{\qz}{%
  {\text{\color{orange}\cmark}}%
}
\newrobustcmd{\jz}{%
  {\text{\color{red}\xmark}}%
}

\newrobustcmd{\projmatrix}[2][placeholder]{%
  {\tensor*{M}{_{#1}^{#2}}}
}
\newrobustcmd{\projorthhum}[2][placeholder]{%
  {\tensor[^#2]{\smash{\check{\mathcal{  P}}}}{#1}}
}
\newrobustcmd{\projorthhumu}[2][placeholder]{%
  {\tensor[^#2]{\smash{\check{\mathcal{  P}}}}{#1}}
}
\newrobustcmd{\projorth}[2][placeholder]{%
  {\tensor[^#2]{\smash{\hat{\mathcal{  P}}}}{#1}}
}
\newrobustcmd{\projlore}[2][placeholder]{%
  {\tensor[^#2]{\hat{\mathcal{  P}}}{#1}}
}

\newrobustcmd{\gensec}[3][placeholder]{%
  {\tensor*[^#1]{\chi}{^{#2}_{\acu{#3}}}}
}

\newrobustcmd{\glfourr}{%
  {\mathrm{GL}(4,\mathbb{R})}%
}

\newrobustcmd{\sltwoc}{%
  {\mathrm{SL}(2,\mathbb{C})}%
}

\newrobustcmd{\poincare}{%
  {\mathbb{R}^{1,3}\rtimes\mathrm{SO}^+(1,3)}%
}

\newrobustcmd{\poincaref}{%
  {\mathrm{P}(1,3)}%
}

\newrobustcmd{\weyl}{%
  {\mathrm{W}(1,3)}%
}

\newrobustcmd{\conformal}{%
  {\mathrm{C}(1,3)}%
}

\newrobustcmd{\diffeomorphism}{%
  {\mathbb{R}^{1,3}}%
}

\newrobustcmd{\soonethree}{%
  {\mathrm{SO}^+(1,3)}%
}

\newrobustcmd{\othree}{%
  {\mathrm{SO}(3)}%
}

\newrobustcmd{\sothree}{%
  {\mathrm{SO}(3)}%
}

\newrobustcmd{\sotwo}{%
  {\mathrm{SO}(2)}%
}

\newrobustcmd{\suthreec}{%
  {\mathrm{SU}(3)_{\text{c}}}%
}

\newrobustcmd{\sutwol}{%
  {\mathrm{SU}(2)_{\text{L}}}%
}

\newrobustcmd{\uoney}{%
  {\mathrm{U}(1)_{\text{Y}}}%
}

\newrobustcmd{\uone}{%
  {\mathrm{U}(1)}%
}

\newrobustcmd{\uoneem}{%
  {\mathrm{U}(1)_{\text{em}}}%
}

\newrobustcmd{\sutwo}{%
  {\mathrm{SU}(2)}%
}

\newrobustcmd{\eplus}{%
  {\tensor{\mathsf{e}}{_{+}}}%
}

\newrobustcmd{\esf}[1]{%
  {\tensor{\mathsf{e}}{_{#1}}}
}%
\newrobustcmd{\esfu}[1]{%
  {\tensor{\mathsf{e}}{^{#1}}}
}%
\newrobustcmd{\gam}[1]{%
  {\tensor{\gamma}{_{#1}}}
}%
\newrobustcmd{\gamu}[1]{%
  {\tensor{\gamma}{^{#1}}}
}%

\newrobustcmd{\planck}{%
  {m_{\text{p}}}%
}

\newrobustcmd{\Pg}{%
	{\Phi_{\text{Nt}}}%
}
\newrobustcmd{\Rh}{%
	{r_{\text{S}}}%
}
\newrobustcmd{\onshell}{%
	{\ =\ }%
}
\newrobustcmd{\Hl}{%
	{h_{\gamma}}%
}
\newrobustcmd{\Kl}{%
	{k_{\gamma}}%
}
\newrobustcmd{\Rb}{%
	{\rho_{\text{Br}}}%
}
\newrobustcmd{\Amond}{%
	{a_{0}}%
}
\newrobustcmd{\Anew}{%
	{a_{\text{Nt}}}%
}
\newrobustcmd{\Aobs}{%
	{a_{\text{Ob}}}%
}
\newrobustcmd{\Rl}{%
	{r_{\gamma}}%
}
\newrobustcmd{\Rp}{%
	{r_{+}}%
}
\newrobustcmd{\Rm}{%
	{r_{-}}%
}
\newrobustcmd{\Risco}{%
	{r_{\pm}}%
}
\newrobustcmd{\Rg}{%
	{r_{\text{Sz}}}%
}
\newrobustcmd{\Kb}{%
	{K_{\text{B}}}%
}

\newrobustcmd{\caligR}{%
  {\mathcal{R}}%
}
\newrobustcmd{\caligT}{%
  {\mathcal{T}}%
}

\newrobustcmd{\pgt}{%
  PGT\textsuperscript{q,+}\ %
}

\newrobustcmd{\unl}[1]{%
  {\mathfrak{#1}}%
}
\newrobustcmd{\ovl}[1]{%
\overline{#1}%
}
\newrobustcmd{\acu}[1]{%
\acute{#1}%
}

\newrobustcmd{\indiq}[2][placeholder]{%
\IfEqCase{#1}{%
  {placeholder}{%
    \IfEqCase{#2}{%
      {1}{\ovl{k}}%
      {2}{\ovl{kl}}%
      {3}{\ovl{klm}}%
    }%
  }%
}[#1]%
}%

\newrobustcmd{\indaq}[2][placeholder]{%
\IfEqCase{#1}{%
  {placeholder}{%
    \IfEqCase{#2}{%
      {1}{\overline{k}}%
      {2}{\overline{kl}}%
      {3}{\overline{klm}}%
    }%
  }%
}[#1]%
}%

\newrobustcmd{\indeq}[2][placeholder]{%
\IfEqCase{#1}{%
  {placeholder}{%
    \IfEqCase{#2}{%
      {1}{k}%
      {2}{kl}%
      {3}{klm}%
    }%
  }%
}[#1]%
}%

\newrobustcmd{\indoq}[2][placeholder]{%
\IfEqCase{#1}{%
  {placeholder}{%
    \IfEqCase{#2}{%
      {1}{\alpha}%
      {2}{\alpha\beta}%
      {3}{\alpha\beta\gamma}%
    }%
  }%
}[#1]%
}%

\newrobustcmd{\fcphi}[1]{%
  \tensor[^{\text{FC}}]{\phi}{_{#1}}%
}

\newrobustcmd{\scphi}[1]{%
  \tensor[^{\text{SC}}]{\phi}{_{#1}}%
}

\newrobustcmd{\fcmul}[1]{%
  \tensor[^{\text{FC}}]{\upsilon}{_{#1}}%
}

\newrobustcmd{\arb}{%
  {\tensor{f}{_{\text{lin}}}}%
}

\newrobustcmd{\scmul}[1]{%
  \tensor[^{\text{SC}}]{\upsilon}{_{#1}}%
}

\newrobustcmd{\foli}[1]{%
\tensor{n}{_{#1}}%
}

\newrobustcmd{\foliu}[1]{%
\tensor{n}{^{#1}}%
}

\newrobustcmd{\covderl}[1]{%
\tensor{\mathcal{D}}{^{\flat}_{\indiq[#1]{1}}}%
}

\newrobustcmd{\covder}[1]{%
\tensor{\mathcal{D}}{_{\indiq[#1]{1}}}%
}
\newrobustcmd{\coder}[1]{%
\tensor{D}{_{\indiq[#1]{1}}}%
}

\newrobustcmd{\deltal}[2]{%
  \tensor*{\delta}{_{\phantom{\flat}}^{\flat}_{#1}^{#2}}%
}

\newrobustcmd{\deltaud}[2]{%
  \tensor*{\delta}{^{#1}_{#2}}%
}
\newrobustcmd{\etau}[1]{%
\tensor{\eta}{^{\indiq[#1]{2}}}%
}
\newrobustcmd{\etaul}[1]{%
\tensor{\eta}{^{\flat}^{\indiq[#1]{2}}}%
}
\newrobustcmd{\etad}[1]{%
\tensor{\eta}{_{\indiq[#1]{2}}}%
}
\newrobustcmd{\etadl}[1]{%
\tensor{\eta}{^{\flat}_{\indiq[#1]{2}}}%
}

\newrobustcmd{\epsul}[1]{%
\tensor{\epsilon}{^{\flat}^{\indiq[#1]{3}}^{\perp}}
}
\newrobustcmd{\epsdl}[1]{%
\tensor{\epsilon}{^{\flat}_{\indiq[#1]{3}}_{\perp}}
}
\newrobustcmd{\epsd}[1]{%
\tensor{\epsilon}{_{\indiq[#1]{3}}_{\perp}}
}
\newrobustcmd{\epsu}[1]{%
\tensor{\epsilon}{^{\indiq[#1]{3}}^{\perp}}
}

\newrobustcmd{\hfl}[2]{%
  \tensor{h}{^{\flat}_{#1}^{#2}}
}

\newrobustcmd{\cgalp}{\tensor{\alpha}{_{\text{CG}}}}

\newrobustcmd{\cbet}[1]{%
  \tensor{\bar{\beta}}{_{#1}}
}
\newrobustcmd{\calp}[1]{%
  \tensor{\bar{\alpha}}{_{#1}}
}

\newrobustcmd{\alpg}[1]{%
  \tensor{\check{\alpha}}{_{#1}}
}
\newrobustcmd{\betg}[1]{%
  \tensor{\check{\beta}}{_{#1}}
}
\newrobustcmd{\calpg}[1]{%
  \tensor{\acu{\alpha}}{_{#1}}
}
\newrobustcmd{\cbetg}[1]{%
  \tensor{\acu{\beta}}{_{#1}}
}

\newrobustcmd{\hub}{%
  {\underline{\mathsf{h}}}
}
\newrobustcmd{\hubm}{%
  {\underline{\mathsf{h}}^{-1}}
}
\newrobustcmd{\hob}{%
  {\bar{\mathsf{h}}}
}
\newrobustcmd{\hobm}{%
  {\bar{\mathsf{h}}^{-1}}
}
\newrobustcmd{\hdet}{%
  {\det \mathsf{h}}
}
\newrobustcmd{\hmdet}{%
  {\det \mathsf{h}^{-1}}
}
\newrobustcmd{\Rsf}{%
  {\mathsf{R}}
}

\newrobustcmd{\alpm}[2][placeholder]{%
  {\tensor*{\hat{\alpha}}{_{#1}^{#2}}}
}
\newrobustcmd{\calpm}[2][placeholder]{%
  \tensor*{\bar{\alpha}}{_{#1}^{#2}}
}

\newrobustcmd{\betm}[2][placeholder]{%
  {\tensor*{\hat{\beta}}{_{#1}^{#2}}}
}
\newrobustcmd{\cbetm}[2][placeholder]{%
  \tensor*{\bar{\beta}}{_{#1}^{#2}}
}

\newrobustcmd{\lamr}{%
  {\zeta_{\mathcal{  R}} }
}
\newrobustcmd{\barlamr}{%
  {\bar{\zeta}_{\mathcal{  R}} }
}
\newrobustcmd{\lamt}{%
  {\zeta_{\mathcal{  T}} }
}
\newrobustcmd{\barlamt}{%
  {\bar{\zeta}_{\mathcal{  T}} }
}

\newrobustcmd{\atmp}[1]{%
  \tensor{\hat{a}}{_{#1}}
}
\newrobustcmd{\btmp}[1]{%
  \tensor{b}{_{#1}}
}

\newrobustcmd{\ctmp}[2][placeholder]{%
  {\tensor*{c}{_{#1}^{#2}}}
}
\newrobustcmd{\dtmp}[2][placeholder]{%
  {\tensor*{d}{_{#1}^{#2}}}
}

\newrobustcmd{\etmp}[1]{%
  \tensor{e}{_{#1}}
}

\newrobustcmd{\batmp}[1]{%
  \tensor{\ovl{a}}{_{#1}}
}
\newrobustcmd{\bbtmp}[1]{%
  \tensor{\ovl{b}}{_{#1}}
}
\newrobustcmd{\bctmp}[1]{%
  \tensor{\ovl{c}}{_{#1}}
}
\newrobustcmd{\bdtmp}[1]{%
  \tensor{\ovl{d}}{_{#1}}
}
\newrobustcmd{\betmp}[1]{%
  \tensor{\ovl{e}}{_{#1}}
}

\newrobustcmd{\ptl}[1]{%
  \tensor{\partial}{#1}
}

\newrobustcmd{\etaf}[1]{%
  \tensor{\eta}{#1}
}

\newrobustcmd{\epsf}[1]{%
  \tensor{\epsilon}{#1}
}

\newrobustcmd{\RSO}[2][placeholder]{%
  {\tensor[^{#2}]{\mathcal{  R}}{#1}}
}
\newrobustcmd{\TSO}[2][placeholder]{%
  {\tensor[^{#2}]{\mathcal{  T}}{#1}}
}
\newrobustcmd{\FSO}[2][placeholder]{%
  {\tensor[^{#2}]{\mathcal{  F}}{#1}}
}
\newrobustcmd{\spinSO}[2][placeholder]{%
  {\tensor[^{#2}]{\sigma}{#1}}
}
\newrobustcmd{\RLambdaSO}[2][placeholder]{%
  {\tensor[^{#2}]{\zeta}{#1}}
}
\newrobustcmd{\TLambdaSO}[2][placeholder]{%
  {\tensor[^{#2}]{\zeta}{#1}}
}
\newrobustcmd{\KSO}[2][placeholder]{%
  {\tensor[^{#2}]{\mathcal{  K}}{#1}}
}

\newrobustcmd{\bper}[2][placeholder]{%
\IfEqCase{#2}{%
  {s}{\tensor{\mathfrak{s}}{#1}}%
  {a}{\tensor{\mathfrak{a}}{#1}}%
  {sbar}{\tensor{\bar{\mathfrak{s}}}{#1}}%
}[\packageError{cosmicclass}{Unidentified Critical Case: #1}{}]%
}

\newrobustcmd{\Jl}{%
  {J^{\flat}}%
}%

\newrobustcmd{\Nl}{%
  {N^{\flat}}%
}%

\newrobustcmd{\haml}[2][placeholder]{%
\IfEqCase{#2}{%
{mom0p}{\tensor{\mathcal{H}}{^{\flat}_{\perp}}}%
{mom1m}{\tensor{\mathcal{H}}{^{\flat}_{\indoq[#1]{1}}}}%
{rot1p}{\tensor{\mathcal{H}}{^{\flat}_{\indaq[#1]{2}}}}%
{rot1m}{\tensor{\mathcal{H}}{^{\flat}_{\perp}_{\indaq[#1]{1}}}}%
}[\packageError{cosmicclass}{Unidentified Critical Case: #1}{}]%
}

\newrobustcmd{\arc}[2][placeholder]{%
\IfEqCase{#2}{%
{B1p}{\tensor{\vartheta}{_{\perp\indiq[#1]{2}}}}%
{B2m}{\tensor[^{\text{T}}]{\vartheta}{_{\indiq[#1]{3}}}}%
{A0m}{\tensor[^{\text{P}}]{\vartheta}{}}%
{A1p}{\tensor{\overset{\wedge}{\vartheta}}{_{\perp\indiq[#1]{2}}}}%
{A1m}{\tensor{\overset{\rightharpoonup}{\vartheta}}{_{\indiq[#1]{1}}}}%
{A2p}{\tensor{\overset{\sim}{\vartheta}}{_{\perp\indiq[#1]{2}}}}%
{A2m}{\tensor[^{\text{T}}]{\vartheta}{_{\perp\indiq[#1]{3}}}}%
}[\packageError{cosmicclass}{Unidentified Critical Case: #1}{}]%
}

\newrobustcmd{\pic}[2][placeholder]{%
\IfEqCase{#2}{%
{B0p}{\varphi}%
{B1p}{\tensor{\overset{\wedge}{\varphi}}{_{\indiq[#1]{2}}}}%
{B1m}{\tensor{\varphi}{_{\perp\indiq[#1]{1}}}}%
{B2p}{\tensor{\overset{\sim}{\varphi}}{_{\indiq[#1]{2}}}}%
{A0p}{\tensor{\varphi}{_\perp}}%
{A0m}{\tensor[^{\text{P}}]{\varphi}{}}%
{A1p}{\tensor{\overset{\wedge}{\varphi}}{_{\perp\indiq[#1]{2}}}}%
{A1m}{\tensor{\overset{\rightharpoonup}{\varphi}}{_{\indiq[#1]{1}}}}%
{A2p}{\tensor{\overset{\sim}{\varphi}}{_{\perp\indiq[#1]{2}}}}%
{A2m}{\tensor[^{\text{T}}]{\varphi}{_{\indiq[#1]{3}}}}%
}[\packageError{cosmicclass}{Unidentified Critical Case: #1}{}]%
}

\newrobustcmd{\picu}[2][placeholder]{%
\IfEqCase{#2}{%
{B0p}{\varphi}%
{B1p}{\tensor{\smash{\overset{\wedge}{\varphi}}}{^{\indiq[#1]{2}}}}%
{B1m}{\tensor{\varphi}{^{\perp\indiq[#1]{1}}}}%
{B2p}{\tensor{\smash{\overset{\sim}{\varphi}}}{^{\indiq[#1]{2}}}}%
{A0p}{\tensor{\varphi}{_\perp}}%
{A0m}{\tensor[^{\text{P}}]{\varphi}{}}%
{A1p}{\tensor{\smash{\overset{\wedge}{\varphi}}}{^{\perp\indiq[#1]{2}}}}%
{A1m}{\tensor{\smash{\overset{\rightharpoonup}{\varphi}}}{^{\indiq[#1]{1}}}}%
{A2p}{\tensor{\smash{\overset{\sim}{\varphi}}}{^{\perp\indiq[#1]{2}}}}%
{A2m}{\tensor[^{\text{T}}]{\varphi}{^{\indiq[#1]{3}}}}%
}[\packageError{cosmicclass}{Unidentified Critical Case: #1}{}]%
}

\newrobustcmd{\picl}[2][placeholder]{%
\IfEqCase{#2}{%
{B0p}{\tensor{\varphi}{^{\flat}}}%
{B1p}{\tensor{\smash{\overset{\wedge}{\varphi}}}{^{\flat}_{\indiq[#1]{2}}}}%
{B1m}{\tensor{\varphi}{^{\flat}_{\perp}_{\indiq[#1]{1}}}}%
{B2p}{\tensor{\smash{\overset{\sim}{\varphi}}}{^{\flat}_{\indiq[#1]{2}}}}%
{A0p}{\tensor{\varphi}{_\perp}^{\flat}}%
{A0m}{\tensor[^{\text{P}}]{\varphi}{^{\flat}}}%
{A1p}{\tensor{\smash{\overset{\wedge}{\varphi}}}{^{\flat}_{\perp\indiq[#1]{2}}}}%
{A1m}{\tensor{\smash{\overset{\rightharpoonup}{\varphi}}}{^{\flat}_{\indiq[#1]{1}}}}%
{A2p}{\tensor{\smash{\overset{\sim}{\varphi}}}{^{\flat}_{\perp\indiq[#1]{2}}}}%
{A2m}{\tensor[^{\text{T}}]{\varphi}{^{\flat}_{\indiq[#1]{3}}}}%
}[\packageError{cosmicclass}{Unidentified Critical Case: #1}{}]%
}

\newrobustcmd{\mull}[2][placeholder]{%
\IfEqCase{#2}{%
{B0p}{\tensor{u}{^{\flat}}}%
{B1p}{\tensor{\smash{\overset{\wedge}{u}}}{^{\flat}_{\indiq[#1]{2}}}}%
{B1m}{\tensor{u}{^{\flat}_{\perp}_{\indiq[#1]{1}}}}%
{B2p}{\tensor{\smash{\overset{\sim}{u}}}{^{\flat}_{\indiq[#1]{2}}}}%
{A0p}{\tensor{u}{_\perp}^{\flat}}%
{A0m}{\tensor[^{\text{P}}]{u}{^{\flat}}}%
{A1p}{\tensor{\smash{\overset{\wedge}{u}}}{^{\flat}_{\perp\indiq[#1]{2}}}}%
{A1m}{\tensor{\smash{\overset{\rightharpoonup}{u}}}{^{\flat}_{\indiq[#1]{1}}}}%
{A2p}{\tensor{\smash{\overset{\sim}{u}}}{^{\flat}_{\perp\indiq[#1]{2}}}}%
{A2m}{\tensor[^{\text{T}}]{u}{^{\flat}_{\indiq[#1]{3}}}}%
}[\packageError{cosmicclass}{Unidentified Critical Case: #1}{}]%
}

\newrobustcmd{\mul}[2][placeholder]{%
\IfEqCase{#2}{%
{B0p}{\tensor{u}{}}%
{B1p}{\tensor{\smash{\overset{\wedge}{u}}}{_{\indiq[#1]{2}}}}%
{B1m}{\tensor{u}{_{\perp}_{\indiq[#1]{1}}}}%
{B2p}{\tensor{\smash{\overset{\sim}{u}}}{_{\indiq[#1]{2}}}}%
{A0p}{\tensor{u}{_\perp}}%
{A0m}{\tensor[^{\text{P}}]{u}{}}%
{A1p}{\tensor{\smash{\overset{\wedge}{u}}}{_{\perp\indiq[#1]{2}}}}%
{A1m}{\tensor{\smash{\overset{\rightharpoonup}{u}}}{_{\indiq[#1]{1}}}}%
{A2p}{\tensor{\smash{\overset{\sim}{u}}}{_{\perp\indiq[#1]{2}}}}%
{A2m}{\tensor[^{\text{T}}]{u}{_{\indiq[#1]{3}}}}%
}[\packageError{cosmicclass}{Unidentified Critical Case: #1}{}]%
}

\newrobustcmd{\PiP}[2][placeholder]{%
\IfEqCase{#2}{%
{B0p}{\hat{\pi}}%
{B1p}{\tensor{\overset{\wedge}{\hat{\pi}}}{_{\indiq[#1]{2}}}}%
{B1m}{\tensor{\hat{\pi}}{_{\perp\indiq[#1]{1}}}}%
{B2p}{\tensor{\overset{\sim}{\hat{\pi}}}{_{\indiq[#1]{2}}}}%
{A0p}{\tensor{\hat{\pi}}{_\perp}}%
{A0m}{\tensor[^{\text{P}}]{\hat{\pi}}{}}%
{A1p}{\tensor{\overset{\wedge}{\hat{\pi}}}{_{\perp\indiq[#1]{2}}}}%
{A1m}{\tensor{\overset{\rightharpoonup}{\hat{\pi}}}{_{\indiq[#1]{1}}}}%
{A2p}{\tensor{\overset{\sim}{\hat{\pi}}}{_{\perp\indiq[#1]{2}}}}%
{A2m}{\tensor[^{\text{T}}]{\hat{\pi}}{_{\indiq[#1]{3}}}}%
}[\packageError{cosmicclass}{Unidentified Critical Case: #1}{}]%
}

\newrobustcmd{\PiPu}[2][placeholder]{%
\IfEqCase{#2}{%
{B0p}{\hat{\pi}}%
{B1p}{\tensor{\smash{\overset{\wedge}{\hat{\pi}}}}{^{\indiq[#1]{2}}}}%
{B1m}{\tensor{\smash{\hat{\pi}}}{^{\perp\indiq[#1]{1}}}}%
{B2p}{\tensor{\smash{\overset{\sim}{\hat{\pi}}}}{^{\indiq[#1]{2}}}}%
{A0p}{\tensor{\smash{\hat{\pi}}}{^\perp}}%
{A0m}{\tensor[^{\text{P}}]{\smash{\hat{\pi}}}{}}%
{A1p}{\tensor{\smash{\overset{\wedge}{\hat{\pi}}}}{^{\perp\indiq[#1]{2}}}}%
{A1m}{\tensor{\smash{\overset{\rightharpoonup}{\hat{\pi}}}}{^{\indiq[#1]{1}}}}%
{A2p}{\tensor{\smash{\overset{\sim}{\hat{\pi}}}}{^{\perp\indiq[#1]{2}}}}%
{A2m}{\tensor[^{\text{T}}]{\smash{\hat{\pi}}}{^{\indiq[#1]{3}}}}%
}[\packageError{cosmicclass}{Unidentified Critical Case: #1}{}]%
}

\newrobustcmd{\sicl}[2][placeholder]{%
\IfEqCase{#2}{%
{B0p}{\tensor{\chi}{^{\flat}}}%
{B1p}{\tensor{\smash{\overset{\wedge}{\chi}}}{^{\flat}_{\indiq[#1]{2}}}}%
{B1m}{\tensor{\chi}{^{\flat}_{\perp}_{\indiq[#1]{1}}}}%
{B2p}{\tensor{\smash{\overset{\sim}{\chi}}}{^{\flat}_{\indiq[#1]{2}}}}%
{A0p}{\tensor{\chi}{^{\flat}_\perp}}%
{A0m}{\tensor[^{\text{P}}]{\chi}{^{\flat}}}%
{A1p}{\tensor{\smash{\overset{\wedge}{\chi}}}{^{\flat}_{\perp\indiq[#1]{2}}}}%
{A1m}{\tensor{\smash{\overset{\rightharpoonup}{\chi}}}{^{\flat}_{\indiq[#1]{1}}}}%
{A2p}{\tensor{\smash{\overset{\sim}{\chi}}}{^{\flat}_{\perp\indiq[#1]{2}}}}%
{A2m}{\tensor[^{\text{T}}]{\chi}{^{\flat}_{\indiq[#1]{3}}}}%
}[\packageError{cosmicclass}{Unidentified Critical Case: #1}{}]%
}

\newrobustcmd{\ticl}[2][placeholder]{%
\IfEqCase{#2}{%
{B0p}{\tensor{\zeta}{^{\flat}}}%
{B1p}{\tensor{\smash{\overset{\wedge}{\zeta}}}{^{\flat}_{\indiq[#1]{2}}}}%
{B1m}{\tensor{\zeta}{^{\flat}_{\perp}_{\indiq[#1]{1}}}}%
{B2p}{\tensor{\smash{\overset{\sim}{\zeta}}}{^{\flat}_{\indiq[#1]{2}}}}%
{A0p}{\tensor{\zeta}{^{\flat}_\perp}}%
{A0m}{\tensor[^{\text{P}}]{\zeta}{^{\flat}}}%
{A1p}{\tensor{\smash{\overset{\wedge}{\zeta}}}{^{\flat}_{\perp\indiq[#1]{2}}}}%
{A1m}{\tensor{\smash{\overset{\rightharpoonup}{\zeta}}}{^{\flat}_{\indiq[#1]{1}}}}%
{A2p}{\tensor{\smash{\overset{\sim}{\zeta}}}{^{\flat}_{\perp\indiq[#1]{2}}}}%
{A2m}{\tensor[^{\text{T}}]{\zeta}{^{\flat}_{\indiq[#1]{3}}}}%
}[\packageError{cosmicclass}{Unidentified Critical Case: #1}{}]%
}

\newrobustcmd{\PiPl}[2][placeholder]{%
\IfEqCase{#2}{%
{B0p}{\tensor{\hat{\pi}}{^{\flat}}}%
{B1p}{\tensor{\smash{\overset{\wedge}{\hat{\pi}}}}{^{\flat}_{\indiq[#1]{2}}}}%
{B1m}{\tensor{\hat{\pi}}{^{\flat}_{\perp}_{\indiq[#1]{1}}}}%
{B2p}{\tensor{\smash{\overset{\sim}{\hat{\pi}}}}{^{\flat}_{\indiq[#1]{2}}}}%
{A0p}{\tensor{\hat{\pi}}{_\perp}^{\flat}}%
{A0m}{\tensor[^{\text{P}}]{\hat{\pi}}{^{\flat}}}%
{A1p}{\tensor{\smash{\overset{\wedge}{\hat{\pi}}}}{^{\flat}_{\perp\indiq[#1]{2}}}}%
{A1m}{\tensor{\smash{\overset{\rightharpoonup}{\hat{\pi}}}}{^{\flat}_{\indiq[#1]{1}}}}%
{A2p}{\tensor{\smash{\overset{\sim}{\hat{\pi}}}}{^{\flat}_{\perp\indiq[#1]{2}}}}%
{A2m}{\tensor[^{\text{T}}]{\hat{\pi}}{^{\flat}_{\indiq[#1]{3}}}}%
}[\packageError{cosmicclass}{Unidentified Critical Case: #1}{}]%
}

\newrobustcmd{\sic}[2][placeholder]{%
\IfEqCase{#2}{%
{B0p}{\chi}%
{B1p}{\tensor{\overset{\wedge}{\chi}}{_{\indiq[#1]{2}}}}%
{B1m}{\tensor{\chi}{_{\perp\indiq[#1]{1}}}}%
{B2p}{\tensor{\overset{\sim}{\chi}}{_{\indiq[#1]{2}}}}%
{A0p}{\tensor{\chi}{_\perp}}%
{A0m}{\tensor[^{\text{P}}]{\chi}{}}%
{A1p}{\tensor{\overset{\wedge}{\chi}}{_{\perp\indiq[#1]{2}}}}%
{A1m}{\tensor{\overset{\rightharpoonup}{\chi}}{_{\indiq[#1]{1}}}}%
{A2p}{\tensor{\overset{\sim}{\chi}}{_{\perp\indiq[#1]{2}}}}%
{A2m}{\tensor[^{\text{T}}]{\chi}{_{\indiq[#1]{3}}}}%
}[\packageError{cosmicclass}{Unidentified Critical Case: #1}{}]%
}

\newrobustcmd{\lorsicpar}[2][placeholder]{%
\IfEqCase{#2}{%
{B0m}{\tensor*[^{\text{P}}]{\smash{\underline{\chi}}}{^{\parallel}}}%
{B1p}{\tensor*{\smash{\overset{\wedge}{\chi}}}{^{\parallel}_{\indiq[#1]{2}}}}%
{B1m}{\tensor*{\chi}{^{\parallel}_{\perp\indiq[#1]{1}}}}%
{B2m}{\tensor*[^{\text{T}}]{\smash{\underline{\chi}}}{^{\parallel}_{\indiq[#1]{3}}}}%
{A0p}{\tensor*{\chi}{^{\parallel}_\perp}}%
{A0m}{\tensor*[^{\text{P}}]{\chi}{^{\parallel}}}%
{A1p}{\tensor*{\smash{\overset{\wedge}{\chi}}}{^{\parallel}_{\perp\indiq[#1]{2}}}}%
{A1m}{\tensor*{\smash{\overset{\rightharpoonup}{\chi}}}{^{\parallel}_{\indiq[#1]{1}}}}%
{A2p}{\tensor*{\smash{\overset{\sim}{\chi}}}{^{\parallel}_{\perp\indiq[#1]{2}}}}%
{A2m}{\tensor*[^{\text{T}}]{\chi}{^{\parallel}_{\indiq[#1]{3}}}}%
}[\packageError{cosmicclass}{Unidentified Critical Case: #1}{}]%
}

\newrobustcmd{\lorsicpir}[2][placeholder]{%
\IfEqCase{#2}{%
{B0p}{\tensor*{\chi}{^{\vDash}}}%
{B1p}{\tensor*{\smash{\overset{\wedge}{\chi}}}{^{\vDash}_{\indiq[#1]{2}}}}%
{B1m}{\tensor*{\chi}{^{\vDash}_{\perp\indiq[#1]{1}}}}%
{B2p}{\tensor*{\smash{\overset{\sim}{\chi}}}{^{\vDash}_{\indiq[#1]{2}}}}%
{A0p}{\tensor*{\chi}{^{\vDash}_\perp}}%
{A0m}{\tensor*[^{\text{P}}]{\chi}{^{\vDash}}}%
{A1p}{\tensor*{\smash{\overset{\wedge}{\chi}}}{^{\vDash}_{\perp\indiq[#1]{2}}}}%
{A1m}{\tensor*{\smash{\overset{\rightharpoonup}{\chi}}}{^{\vDash}_{\indiq[#1]{1}}}}%
{A2p}{\tensor*{\smash{\overset{\sim}{\chi}}}{^{\vDash}_{\perp\indiq[#1]{2}}}}%
{A2m}{\tensor*[^{\text{T}}]{\chi}{^{\vDash}_{\indiq[#1]{3}}}}%
}[\packageError{cosmicclass}{Unidentified Critical Case: #1}{}]%
}

\newrobustcmd{\lorsicper}[2][placeholder]{%
\IfEqCase{#2}{%
{B0p}{\tensor*{\chi}{^{\perp}}}%
{B1p}{\tensor*{\smash{\overset{\wedge}{\chi}}}{^{\perp}_{\indiq[#1]{2}}}}%
{B1m}{\tensor*{\chi}{^{\perp}_{\perp\indiq[#1]{1}}}}%
{B2p}{\tensor*{\smash{\overset{\sim}{\chi}}}{^{\perp}_{\indiq[#1]{2}}}}%
{A0p}{\tensor*{\chi}{^{\perp}_\perp}}%
{A0m}{\tensor*[^{\text{P}}]{\chi}{^{\perp}}}%
{A1p}{\tensor*{\smash{\overset{\wedge}{\chi}}}{^{\perp}_{\perp\indiq[#1]{2}}}}%
{A1m}{\tensor*{\smash{\overset{\rightharpoonup}{\chi}}}{^{\perp}_{\indiq[#1]{1}}}}%
{A2p}{\tensor*{\smash{\overset{\sim}{\chi}}}{^{\perp}_{\perp\indiq[#1]{2}}}}%
{A2m}{\tensor*[^{\text{T}}]{\chi}{^{\perp}_{\indiq[#1]{3}}}}%
}[\packageError{cosmicclass}{Unidentified Critical Case: #1}{}]%
}

\newrobustcmd{\Tl}[2][placeholder]{%
\IfEqCase{#2}{%
{B0p}{\tensor{\chi}{^{\flat}}}%
{B1p}{\tensor{\smash{\overset{\wedge}{\chi}}}{^{\flat}_{\indiq[#1]{2}}}}%
{B1m}{\tensor{\chi}{^{\flat}_{\perp}_{\indiq[#1]{1}}}}%
{B2p}{\tensor{\smash{\overset{\sim}{\chi}}}{^{\flat}_{\indiq[#1]{2}}}}%
{A0p}{\tensor{\chi}{^{\flat}_\perp}}%
{A0m}{\tensor[^{\text{P}}]{\mathcal{T}}{^{\flat}}}%
{A1p}{\tensor{\smash{\overset{\wedge}{\chi}}}{^{\flat}_{\perp\indiq[#1]{2}}}}%
{A1m}{\tensor{\smash{\overset{\rightharpoonup}{\mathcal{T}}}}{^{\flat}_{\indiq[#1]{1}}}}%
{A2p}{\tensor{\smash{\overset{\sim}{\chi}}}{^{\flat}_{\perp\indiq[#1]{2}}}}%
{A2m}{\tensor[^{\text{T}}]{\mathcal{T}}{^{\flat}_{\indiq[#1]{3}}}}%
}[\tensor{\mathcal{T}}{^{\flat}_{\indiq[#1]{3}}}]%
}

\newrobustcmd{\cT}[2][placeholder]{%
\IfEqCase{#2}{%
{B1p}{\tensor{\mathcal{T}}{_{\perp\indiq[#1]{2}}}}%
{B1m}{\tensor{\overset{\rightharpoonup}{\mathcal{T}}}{_{\indiq[#1]{1}}}}%
{A0m}{\tensor[^{\text{P}}]{\mathcal{T}}{}}%
{A2m}{\tensor[^{\text{T}}]{\mathcal{T}}{_{\indiq[#1]{3}}}}%
}[\packageError{cosmicclass}{Unidentified Critical Case: #1}{}]%
}

\newrobustcmd{\cTLambda}[2][placeholder]{%
\IfEqCase{#2}{%
{B1p}{\tensor{\zeta}{_{\perp\indiq[#1]{2}}}}%
{B1m}{\tensor{\overset{\rightharpoonup}{\zeta}}{_{\indiq[#1]{1}}}}%
{A0m}{\tensor[^{\text{P}}]{\zeta}{}}%
{A2m}{\tensor[^{\text{T}}]{\zeta}{_{\indiq[#1]{3}}}}%
}[\packageError{cosmicclass}{Unidentified Critical Case: #1}{}]%
}

\newrobustcmd{\cTpic}[2][placeholder]{%
\IfEqCase{#2}{%
{B1p}{\tensor{\phi}{_{\perp\indiq[#1]{2}}}}%
{B1m}{\tensor{\overset{\rightharpoonup}{\phi}}{_{\indiq[#1]{1}}}}%
{A0m}{\tensor[^{\text{P}}]{\phi}{}}%
{A2m}{\tensor[^{\text{T}}]{\phi}{_{\indiq[#1]{3}}}}%
}[\packageError{cosmicclass}{Unidentified Critical Case: #1}{}]%
}

\newrobustcmd{\cTpicl}[2][placeholder]{%
\IfEqCase{#2}{%
{B1p}{\tensor{\phi}{^{\flat}_{\perp\indiq[#1]{2}}}}%
{B1m}{\tensor{\overset{\rightharpoonup}{\phi}}{^{\flat}_{\indiq[#1]{1}}}}%
{A0m}{\tensor[^{\text{P}}]{\phi}{^{\flat}}}%
{A2m}{\tensor[^{\text{T}}]{\phi}{^{\flat}_{\indiq[#1]{3}}}}%
}[\packageError{cosmicclass}{Unidentified Critical Case: #1}{}]%
}

\newrobustcmd{\cTPiP}[2][placeholder]{%
\IfEqCase{#2}{%
{B1p}{\tensor{\varpi}{_{\perp\indiq[#1]{2}}}}%
{B1m}{\tensor{\overset{\rightharpoonup}{\varpi}}{_{\indiq[#1]{1}}}}%
{A0m}{\tensor[^{\text{P}}]{\varpi}{}}%
{A2m}{\tensor[^{\text{T}}]{\varpi}{_{\indiq[#1]{3}}}}%
}[\packageError{cosmicclass}{Unidentified Critical Case: #1}{}]%
}

\newrobustcmd{\cTl}[2][placeholder]{%
\IfEqCase{#2}{%
{B1p}{\tensor{\mathcal{T}}{^{\flat}_{\perp\indiq[#1]{2}}}}%
{B1m}{\tensor{\smash{\overset{\rightharpoonup}{\mathcal{T}}}}{^{\flat}_{\indiq[#1]{1}}}}%
{A0m}{\tensor[^{\text{P}}]{\mathcal{T}}{^{\flat}}}%
{A2m}{\tensor[^{\text{T}}]{\mathcal{T}}{^{\flat}_{\indiq[#1]{3}}}}%
}[\packageError{cosmicclass}{Unidentified Critical Case: #1}{}]%
}

\newrobustcmd{\cTu}[2][placeholder]{%
\IfEqCase{#2}{%
{B1p}{\tensor{\mathcal{T}}{^{\perp\indiq[#1]{2}}}}%
{B1m}{\tensor{\smash{\overset{\rightharpoonup}{\mathcal{T}}}}{^{\indiq[#1]{1}}}}%
{A0m}{\tensor[^{\text{P}}]{\mathcal{T}}{}}%
{A2m}{\tensor[^{\text{T}}]{\mathcal{T}}{^{\indiq[#1]{3}}}}%
}[\packageError{cosmicclass}{Unidentified Critical Case: #1}{}]%
}

\newrobustcmd{\ncTLambda}[2][placeholder]{%
\IfEqCase{#2}{%
{B0p}{\tensor{\zeta}{^{\indiq[#1]{1}}_{\indiq[#1]{1}\perp}}}%
{B1p}{\tensor{\zeta}{_{[\indiq[#1]{2}]\perp}}}%
{B1m}{\tensor{\zeta}{_{\perp\indiq[#1]{1}\perp}}}%
{B2p}{\tensor{\zeta}{_{\langle\indiq[#1]{2}\rangle\perp}}}%
}[\packageError{cosmicclass}{Unidentified Critical Case: #1}{}]%
}

\newrobustcmd{\ncTmul}[2][placeholder]{%
\IfEqCase{#2}{%
{B0p}{\tensor{\upsilon}{^{\indiq[#1]{1}}_{\indiq[#1]{1}\perp}}}%
{B1p}{\tensor{\upsilon}{_{[\indiq[#1]{2}]\perp}}}%
{B1m}{\tensor{\upsilon}{_{\perp\indiq[#1]{1}\perp}}}%
{B2p}{\tensor{\upsilon}{_{\langle\indiq[#1]{2}\rangle\perp}}}%
}[\packageError{cosmicclass}{Unidentified Critical Case: #1}{}]%
}

\newrobustcmd{\ncTpic}[2][placeholder]{%
\IfEqCase{#2}{%
{B0p}{\tensor{\phi}{^{\indiq[#1]{1}}_{\indiq[#1]{1}\perp}}}%
{B1p}{\tensor{\phi}{_{[\indiq[#1]{2}]\perp}}}%
{B1m}{\tensor{\phi}{_{\perp\indiq[#1]{1}\perp}}}%
{B2p}{\tensor{\phi}{_{\langle\indiq[#1]{2}\rangle\perp}}}%
}[\packageError{cosmicclass}{Unidentified Critical Case: #1}{}]%
}

\newrobustcmd{\ncTpicl}[2][placeholder]{%
\IfEqCase{#2}{%
  {B0p}{\tensor{\phi}{^{\flat}^{\indiq[#1]{1}}_{\indiq[#1]{1}\perp}}}%
{B1p}{\tensor{\phi}{^{\flat}_{[\indiq[#1]{2}]\perp}}}%
{B1m}{\tensor{\phi}{^{\flat}_{\perp\indiq[#1]{1}\perp}}}%
{B2p}{\tensor{\phi}{^{\flat}_{\langle\indiq[#1]{2}\rangle\perp}}}%
}[\packageError{cosmicclass}{Unidentified Critical Case: #1}{}]%
}

\newrobustcmd{\ncTPiP}[2][placeholder]{%
\IfEqCase{#2}{%
{B0p}{\tensor{\varpi}{^{\indiq[#1]{1}}_{\indiq[#1]{1}\perp}}}%
{B1p}{\tensor{\varpi}{_{[\indiq[#1]{2}]\perp}}}%
{B1m}{\tensor{\varpi}{_{\perp\indiq[#1]{1}\perp}}}%
{B2p}{\tensor{\varpi}{_{\langle\indiq[#1]{2}\rangle\perp}}}%
}[\packageError{cosmicclass}{Unidentified Critical Case: #1}{}]%
}

\newrobustcmd{\ncT}[2][placeholder]{%
\IfEqCase{#2}{%
{B0p}{\tensor{\mathcal{T}}{^{\indiq[#1]{1}}_{\indiq[#1]{1}\perp}}}%
{B1p}{\tensor{\mathcal{T}}{_{[\indiq[#1]{2}]\perp}}}%
{B1m}{\tensor{\mathcal{T}}{_{\perp\indiq[#1]{1}\perp}}}%
{B2p}{\tensor{\mathcal{T}}{_{\langle\indiq[#1]{2}\rangle\perp}}}%
}[\packageError{cosmicclass}{Unidentified Critical Case: #1}{}]%
}

\newrobustcmd{\cR}[2][placeholder]{%
\IfEqCase{#2}{%
{A0p}{\tensor{\underline{\mathcal{R}}}{}}%
{A0m}{\tensor[^{\text{P}}]{\mathcal{R}}{_{\perp\circ}}}%
{A1p}{\tensor{\underline{\mathcal{R}}}{_{[\indiq[#1]{2}]}}}%
{A1m}{\tensor{\mathcal{R}}{_{\perp\indiq[#1]{1}}}}%
{A2p}{\tensor{\underline{\mathcal{R}}}{_{\langle\indiq[#1]{2}\rangle}}}%
{A2m}{\tensor[^{\text{T}}]{\mathcal{R}}{_{\perp\indiq[#1]{3}}}}%
}[\packageError{cosmicclass}{Unidentified Critical Case: #1}{}]%
}

\newrobustcmd{\cRLambda}[2][placeholder]{%
\IfEqCase{#2}{%
{A0p}{\tensor{\underline{\zeta}}{}}%
{A0m}{\tensor[^{\text{P}}]{\zeta}{_{\perp\circ}}}%
{A1p}{\tensor{\underline{\zeta}}{_{[\indiq[#1]{2}]}}}%
{A1m}{\tensor{\zeta}{_{\perp\indiq[#1]{1}}}}%
{A2p}{\tensor{\underline{\zeta}}{_{\langle\indiq[#1]{2}\rangle}}}%
{A2m}{\tensor[^{\text{T}}]{\zeta}{_{\perp\indiq[#1]{3}}}}%
}[\packageError{cosmicclass}{Unidentified Critical Case: #1}{}]%
}

\newrobustcmd{\cRl}[2][placeholder]{%
\IfEqCase{#2}{%
  {A0p}{\tensor{\underline{\mathcal{R}}}{^{\flat}}}%
{A0m}{\tensor[^{\text{P}}]{\mathcal{R}}{^{\flat}_{\perp\circ}}}%
{A1p}{\tensor{\underline{\mathcal{R}}}{^{\flat}_{[\indiq[#1]{2}]}}}%
{A1m}{\tensor{\mathcal{R}}{^{\flat}_{\perp\indiq[#1]{1}}}}%
{A2p}{\tensor{\underline{\mathcal{R}}}{^{\flat}_{\langle\indiq[#1]{2}\rangle}}}%
{A2m}{\tensor[^{\text{T}}]{\mathcal{R}}{^{\flat}_{\perp\indiq[#1]{3}}}}%
}[\packageError{cosmicclass}{Unidentified Critical Case: #1}{}]%
}

\newrobustcmd{\cRu}[2][placeholder]{%
\IfEqCase{#2}{%
{A0p}{\tensor{\underline{\mathcal{R}}}{}}%
{A0m}{\tensor[^{\text{P}}]{\mathcal{R}}{_{\perp\circ}}}%
{A1p}{\tensor{\underline{\mathcal{R}}}{^{[\indiq[#1]{2}]}}}%
{A1m}{\tensor{\mathcal{R}}{^{\perp\indiq[#1]{1}}}}%
{A2p}{\tensor{\underline{\mathcal{R}}}{^{\langle\indiq[#1]{2}\rangle}}}%
{A2m}{\tensor[^{\text{T}}]{\mathcal{R}}{^{\perp\indiq[#1]{3}}}}%
}[\packageError{cosmicclass}{Unidentified Critical Case: #1}{}]%
}

\newrobustcmd{\ncR}[2][placeholder]{%
\IfEqCase{#2}{%
  {A0p}{\tensor{\mathcal{R}}{_{\perp\perp}}}%
{A0m}{\tensor[^{\text{P}}]{\mathcal{R}}{_{\circ\perp}}}%
{A1p}{\tensor{\mathcal{R}}{_{\perp[\indiq[#1]{2}]\perp}}}%
{A1m}{\tensor{\mathcal{R}}{_{\indiq[#1]{1}\perp}}}%
{A2p}{\tensor{\mathcal{R}}{_{\perp\langle\indiq[#1]{2}\rangle\perp}}}%
{A2m}{\tensor[^{\text{T}}]{\mathcal{R}}{_{\indiq[#1]{3}\perp}}}%
}[\packageError{cosmicclass}{Unidentified Critical Case: #1}{}]%
}

\newrobustcmd{\ncRLambda}[2][placeholder]{%
\IfEqCase{#2}{%
  {A0p}{\tensor{\zeta}{_{\perp\perp}}}%
{A0m}{\tensor[^{\text{P}}]{\zeta}{_{\circ\perp}}}%
{A1p}{\tensor{\zeta}{_{\perp[\indiq[#1]{2}]\perp}}}%
{A1m}{\tensor{\zeta}{_{\indiq[#1]{1}\perp}}}%
{A2p}{\tensor{\zeta}{_{\perp\langle\indiq[#1]{2}\rangle\perp}}}%
{A2m}{\tensor[^{\text{T}}]{\zeta}{_{\indiq[#1]{3}\perp}}}%
}[\packageError{cosmicclass}{Unidentified Critical Case: #1}{}]%
}

\newrobustcmd{\Proj}[2][placeholder]{%
\IfEqCase{#2}{%
  {A2m}{\tensor[^{\text{T}}]{\check{\mathcal{P}}}{#1}}%
}[\packageError{cosmicclass}{Unidentified Critical Case: #1}{}]%
}

\newrobustcmd{\Projl}[2][placeholder]{%
\IfEqCase{#2}{%
  {A2m}{\tensor[^{\text{T}}]{\check{\mathcal{P}}}{^{\flat}#1}}%
}[\packageError{cosmicclass}{Unidentified Critical Case: #1}{}]%
}

\newrobustcmd{\fA}{%
  {\tensor{\mathcal{  A}}{_{\acu{u}}}}%
}
\newrobustcmd{\fB}{%
  {\tensor{\mathcal{  B}}{_{\acu{v}}}}%
}
\newrobustcmd{\fC}{%
  {\tensor{\mathcal{  C}}{^{\acu{v}}}}%
}
\newrobustcmd{\fphi}{%
  {\tensor{\phi}{^{\acu{w}}}}%
}
\newrobustcmd{\fpi}{%
  {\tensor{\pi}{_{\acu{w}}}}%
}

\newrobustcmd{\covard}[2]{%
  {\frac{\bar{\delta}#1}{\bar{\delta}#2}}
}
\newrobustcmd{\copard}[2]{%
  {\frac{\bar{\partial}#1}{\bar{\partial}#2}}
}
\newrobustcmd{\pard}[2]{%
  {\frac{\partial #1}{\partial #2}}
}

\newrobustcmd{\PPM}[1]{%
  {\left[\tensor*{\mathsf{M}}{_{\ }^{\left(\text{#1}\right)}}\right]}%
}

\tikzset{
  good/.style={circle, opacity=0.7, draw=green!60, fill=green!5, line width=.8mm, minimum size=3.5mm},
  bad/.style={circle, opacity=0.7, draw=red!60, fill=red!5, line width=.8mm, minimum size=3.5mm},
  badx/.style={circle, opacity=0.4, draw=red!60, fill=red!5, line width=.8mm, minimum size=3.5mm},
  save/.style={circle, opacity=0.7, draw=green!60, fill=green!60, line width=1.5mm, minimum size=3.5mm},
  kill/.style={circle, opacity=0.7, draw=red!60, fill=red!60, line width=1.5mm, minimum size=3.5mm},
  bind/.style={draw=green!60, opacity=0.7, line width=1mm,},
  wrap/.style={draw=red!60, opacity=0.7, line width=1mm,},
}

\DeclareTOCStyleEntry[numwidth=20pt,linefill=\bfseries\TOCLineLeaderFill]{tocline}{section}
\DeclareTOCStyleEntry[entryformat=\textit,numwidth=10pt,linefill=\TOCLineLeaderFill]{tocline}{subsection}
\DeclareTOCStyleEntry[entryformat=\textit,numwidth=10pt,linefill=\TOCLineLeaderFill]{tocline}{subsubsection}

\usepackage{hyperref}
\hypersetup{   colorlinks = true,   linkcolor = Blue,   citecolor = Blue,   filecolor = Blue,   urlcolor = Blue%
   }\usepackage[capitalize]{cleveref}

\usepackage{soul}

\newcommand{\myDel}[1]{{\color[RGB]{0,100,0}\ifmmode\cancel{#1}\else\st{#1}\fi}}

\begin{document}

\title{The radial metric function does not identify null surfaces}

\author{Yi-Hsiung Hsu}
\email{yhh36@cam.ac.uk}
\affiliation{Astrophysics Group, Cavendish Laboratory, JJ Thomson Avenue, Cambridge CB3 0HE, UK}

\author{Will Barker}
\email{barker@fzu.cz}
\affiliation{Astrophysics Group, Cavendish Laboratory, JJ Thomson Avenue, Cambridge CB3 0HE, UK}
\affiliation{Kavli Institute for Cosmology, Madingley Road, Cambridge CB3 0HA, UK}
\affiliation{Central European Institute for Cosmology and Fundamental Physics, Institute of Physics of the Czech Academy of Sciences, Na Slovance 1999/2, 182 00 Prague 8, Czechia}

\author{Michael Hobson}
\email{mph@mrao.cam.ac.uk}
\affiliation{Astrophysics Group, Cavendish Laboratory, JJ Thomson Avenue, Cambridge CB3 0HE, UK}

\author{Anthony Lasenby}
\email{a.n.lasenby@mrao.cam.ac.uk}
\affiliation{Astrophysics Group, Cavendish Laboratory, JJ Thomson Avenue, Cambridge CB3 0HE, UK}
\affiliation{Kavli Institute for Cosmology, Madingley Road, Cambridge CB3 0HA, UK}

\begin{abstract}
	We investigate the conditions under which a hypersurface becomes null through the use of coordinate transformations. We demonstrate that, in static spacetimes, the correct criterion for a surface to be null is~$\tensor{g}{_{tt}} = 0$, rather than~$\tensor{g}{^{rr}} = 0$, in agreement with the results of Vollick. We further show that, if a Kruskal-like coordinate exists, the proxy condition~$\tensor{g}{^{rr}} = 0$ is equivalent to~$\tensor{g}{_{tt}} = 0$ if~$\partial_r\tensor{g}{_{tt}} \neq 0$ and both~$\tensor{g}{^{rr}}$ and~$\tensor{g}{_{tt}}$ vanish at the same rate near the horizon. Our method extends naturally to axisymmetric stationary spacetimes, for which we demonstrate that the condition~$\det\big(\tensor{h}{_{ab}}\big) = 0$ for the induced metric on a null hypersurface is recovered. By contrast with the induced metric approach, our method provides a physical perspective that connects the general null condition with its underlying relationship to photon geodesics.
\end{abstract}
\maketitle
\tableofcontents

\section{Introduction}\label{sec:intro}

A null hypersurface is one for which the normal vector~$\tensor{n}{_\mu}$ is null,~$\tensor{n}{_\mu}\tensor{n}{^\mu}=0$~\cite{Wald:1984rg}. For a hypersurface that can be defined by~$f(r) = 0$, where~$f$ is some function of a spacelike (typically radial) coordinate~$r$, a vanishing radial inverse metric component~$\tensor{g}{^{rr}}=0$ is often used as a proxy condition to identify the surface as null~\cite{carroll2004spacetime}. This condition can be readily motivated: if one considers the normal vector~$\tensor{n}{_\mu} = \tensor{\nabla}{_\mu} f = \tensor{\partial}{_\mu} f$ to the surface in some coordinates~$[\tensor{x}{^\mu}] = [t,r,\theta,\phi]$, say, then its components are~$[\tensor{n}{_\mu}]=[0,\tensor{\partial}{_r} f,0,0]$ and its norm is simply~$\tensor{n}{_\mu}\tensor{n}{^\mu}=\tensor{g}{^{rr}}(\partial_r f)^2$. Thus, the condition~$\tensor{g}{^{rr}} = 0$ appears to indicate that the surface is null. 

By contrast, for a static spacetime in a coordinate system for which the metric takes the form~$\mathrm{d}s^2 = \tensor{g}{_{tt}}\,\mathrm{d}t^2 + \tensor{g}{_{ij}}\,\mathrm{d}\tensor{x}{^i}\,\mathrm{d}\tensor{x}{^j}$, with~$\tensor{g}{_{tt}}$ and~$\tensor{g}{_{ij}}$ depending on the spatial coordinates~$\tensor{x}{^i}$ but not the time~$t\equiv \tensor{x}{^0}$, Vishveshwara~\cite{10.1063/1.1664717} proposed that null surfaces should instead be identified by the condition~$\tensor{\mathbf{e}}{_t}\cdot\tensor{\mathbf{e}}{_t} \equiv \tensor{g}{_{tt}} = 0$, where~$\tensor{\mathbf{e}}{_t}$ is the~$t$-coordinate basis vector, which is timelike and hypersurface orthogonal in the static region on one side of the surface. Vishveshwara arrived at this conclusion by considering the normal to the surface of constant~$\tensor{\mathbf{e}}{_t}\cdot\tensor{\mathbf{e}}{_t}$. However, Vollick~\cite{Vollick:2015aa} later identified a loophole in Vishveshwara’s argument, and demonstrated that, in fact, the condition for the normal to the surface of constant~$\tensor{\mathbf{e}}{_t}\cdot\tensor{\mathbf{e}}{_t}$ to be null is
\begin{align}\label{eq:vollick}
	\tensor{g}{^{ij}}\tensor{\partial}{_i}\tensor{g}{_{tt}}\tensor{\partial}{_j}\tensor{g}{_{tt}} = 0.
\end{align} 
For the special case of a spherically symmetric static spacetime with metric~$\mathrm{d}s^2 = \tensor{g}{_{tt}}(r)\,\mathrm{d}t^2 + \tensor{g}{_{rr}}(r)\,\mathrm{d}r^2 + r^2\,\mathrm{d}\Omega^2$, and~$\tensor{\partial}{_r}\tensor{g}{_{tt}} \neq 0$, the condition~\cref{eq:vollick} reduces to~$\tensor{g}{^{rr}}=0$, in agreement with the analysis above.

By its very nature, however, the condition~$\tensor{g}{^{rr}} = 0$ can be problematic. In particular, for spacetimes in which~$\tensor{g}{_{rr}} = 1/\tensor{g}{^{rr}}$, the coordinates become singular at the surface of interest, and so the condition~$\tensor{g}{^{rr}} = 0$ may be a coordinate-dependent artifact arising from the fact that, near the surface,~$r$ is not related in a finite manner to a proper radial distance. More generally, as we will see, the condition~\cref{eq:vollick} holds only in nonsingular coordinate systems, where the metric components remain finite and nonzero, and so may be inapplicable in some singular situations. 

Somewhat ironically, despite identifying a loophole in Vishveshwara's argument, Vollick's own analysis, which considers both non-singular coordinate systems and examination of the induced metric on the surface, recovers~$\tensor{g}{_{tt}} = 0$ as the appropriate condition to identify null surfaces in spherically symmetric static spacetimes. 

Indeed, the most general condition for a hypersurface to be null in any spacetime is that the determinant of its induced metric vanishes, i.e.,~$\det\big(\tensor{h}{_{ab}}\big) = 0$, where 
\begin{align}
	\tensor{h}{_{ab}}\equiv\tensor{g}{_{\mu\nu}}\frac{\partial\tensor{x}{^\mu}}{\partial\tensor{y}{^a}}\frac{\partial\tensor{x}{^\nu}}{\partial\tensor{y}{^b}} \equiv \tensor{g}{_{\mu\nu}}\tensor{e}{_a^\mu}\tensor{e}{_b^\nu},
\end{align}
in which~$\tensor{y}{^a}$ are a set of coordinates that parameterise the surface and we have defined the quantities~$\tensor{e}{_a^\mu}$. The condition~$\det\big(\tensor{h}{_{ab}}\big) = 0$ arises from the fact that for a null hypersurface the normal vector is also a tangent vector, and hence expressible as~$\tensor{n}{^\mu}=\tensor{v}{^a}\tensor{e}{_a^\mu}$. Contracting~$\tensor{v}{^b}$ with the induced metric, we obtain~$\tensor{h}{_{ab}} \tensor{v}{^b} \onshell \tensor{g}{_{\mu\nu}} \left(\tensor{n}{^\nu} \tensor{e}{_a^\mu}\right) \onshell 0$, which shows that there exists a nontrivial direction on the hypersurface along which the induced metric vanishes, since~$\tensor{n}{^\mu}$ is a non-zero tangent vector with zero length. For a nontrivial~$\tensor{v}{^b}$, the above relation is only solvable if~$\det\big(\tensor{h}{_{ab}}\big) = 0$, which occurs since the intrinsic null direction in the tangent space reduces the rank of the induced metric by one, rendering it degenerate. While this condition is general and hence practically useful, its relation to photon trajectories may not be immediately apparent.

In this letter, we therefore analyse null surfaces by instead performing coordinate transformations, in order to explore the relationships between the aforementioned proxy conditions and their physical interpretation. Our focus will be on coordinates similar to those encountered in the Eddington--Finkelstein (EF) and Kruskal schemes, and we discuss these in~\cref{sec:2d}. While EF-like coordinates are sufficient to identify null surfaces, Kruskal-like coordinates provide a regular description of the surface itself, thus ensuring that the metric remains finite and nonzero, and that the surface is located at a finite coordinate value. We first demonstrate these approaches for spherically symmetric static spacetimes in~\cref{sec:2d}, for which we recover the condition~$\tensor{g}{_{tt}}=0$. We then show in~\cref{sec:GeneralS} that our analysis naturally extends to general static spacetimes, for which we arrive at the same condition, but also show that one recovers~\cref{eq:vollick} if~$\tensor{\partial}{_r}\tensor{g}{_{tt}} \neq 0$ and the ratio~$\tensor{g}{_{tt}}\tensor{g}{^{rr}}$ remains finite and nonzero in the vicinity of the surface of interest. We further show that our method also accommodates axisymmetric stationary spacetimes in~\cref{sec:stationary}, for which we demonstrate that one recovers the condition~$\mbox{det}\big(h_{ab}\big) = 0$ on the induced metric,
and we suggest how our approach may be extended to more general spacetimes in~\cref{sec:generalisations}. We apply our method to several wormhole examples in~\cref{sec:example}, and present our conclusions in~\cref{sec:conclusion}. We use the `West Coast' or `mostly minus' metric signature. Starting in the middle of both alphabets, Greek coordinate indices~$\mu$,~$\nu$, etc., run from zero to three, and Roman coordinate indices~$i$,~$j$, etc., are spatial, running from one to three. Roman indices~$a$,~$b$, etc., from the start of the alphabet, refer to adapted coordinates with respect to any other hypersurface.

\section{Theoretical development}\label{sec:NS}
We first investigate the existence of null surfaces by performing coordinate transformations to EF-like and Kruskal-like coordinates. This procedure consists of two steps: (i) defining the tortoise coordinate through an appropriate transformation, and (ii) transforming into EF-like or Kruskal-like coordinates, if feasible. 

\subsection{Spherically symmetric static spacetime}\label{sec:2d}
\subsubsection{Schwarzschild-like and tortoise coordinates}\label{sec:tortoise}
The spherically symmetric static spacetime provides the simplest case for analysis, since it evidently reduces to a two-dimensional problem. The general metric in Schwarzschild-like coordinates is given by 
\begin{align}\label{eq:metric}
	\mathrm{d}s^2 \onshell \TimeFunc(r)\ \mathrm{d}t^2 - \SpaceFunc(r)\ \mathrm{d}r^2 - r^2 \mathrm{d}\Omega^2.
\end{align} 
To introduce the tortoise coordinate~$\Bar{r}$, we consider a radial null curve by setting~$\mathrm{d}s = 0$ and~$\mathrm{d}\Omega = 0$, which yields 
\begin{align}\label{eq:tortoise_prime}
  \frac{\mathrm{d}t}{\mathrm{d}r} = \sqrt{\frac{\SpaceFunc}{\TimeFunc}} \equiv \frac{\mathrm{d}\Bar{r}}{\mathrm{d}r}.
\end{align} 
By expressing~\cref{eq:metric} in terms of~$\Bar{r}$, we obtain 
\begin{align}\label{eq:ds_rbar}
  \mathrm{d}s^2 \onshell \TimeFunc(r) \left( \mathrm{d}t^2 - \mathrm{d}\Bar{r}^2 \right),
\end{align} 
and thus~\cref{eq:ds_rbar} defines the tortoise coordinate system.

Consider first the case where~$\TimeFunc\big(\Rh\big)$ is finite and nonzero, where $r=\Rh$ defines the surface of interest. Then, the radial null curves are diagonal lines in the plane of~$t$ and~$\Bar{r}$, whilst the surface is a vertical line. Therefore, a surface where~$\TimeFunc\big(\Rh\big)$ is finite and nonzero is not a null surface. In~\cref{sec:EFT}, we will show that the opposite case with~$\TimeFunc(\Rh) = 0$ is a null surface. Moreover, if the behaviour of~$\SpaceFunc$ and~$\Bar{r}$ on the surface satisfy certain conditions shown in~\cref{sec:KT}, we will recover~\cref{eq:vollick}.

\subsubsection{Eddington--Finkelstein-like coordinates}\label{sec:EFT}
If~$\TimeFunc(\Rh) = 0$ then the metric is singular at the surface: an additional transformation is required, beyond that which took us from~\cref{eq:metric} to~\cref{eq:ds_rbar}, to determine whether the surface is null. The EF-like transformation can provide a coordinate system in which some metric components remain finite and regular on the surface, allowing for a clearer diagnosis.

Defining~$v \equiv t - \Bar{r}$ and~$\mathrm{d}\Tilde{r} \equiv \TimeFunc \mathrm{d}\Bar{r}$, we can rewrite~\cref{eq:ds_rbar} as 
\begin{align} \label{eq:EF_metic}
	\mathrm{d}s^2 \onshell \TimeFunc(r)\ \mathrm{d}v^2 + 2 \mathrm{d}v\ \mathrm{d} \Tilde{r}.
\end{align} 
Since~$v$ depends on~$t$, it is generally not a constant, but~$\Tilde{r}$ depends only on~$\Bar{r}$, i.e. on~$r$, and so the surface of constant~$r=\Rh$ corresponds to a surface of constant~$\Tilde{r}$. In the limit~$r \mapsto \Rh$, as~$\TimeFunc(r)\mapsto 0$,~\cref{eq:EF_metic} simplifies to~$\mathrm{d}s^2 \onshell 2 \mathrm{d}v\ \mathrm{d} \Tilde{r}$, so the surface is once again a diagonal line in the spacetime diagram, and hence null.

\subsubsection{Kruskal-like coordinates}\label{sec:KT}
We further analyse~\cref{eq:ds_rbar} by introducing Kruskal-like coordinates. Again we consider the case where~$\TimeFunc(\Rh) = 0$. Since the new chart should be regular at the surface, we introduce the new lightcone coordinates as 
\begin{subequations}
  \begin{align}
	  V &\equiv e^{t/2\Rh} \mathscr{F}\left(\TimeFunc\right), \label{eq:V}\\
	  U &\equiv -e^{-t/2\Rh} \mathscr{F}\left(\TimeFunc\right), \label{eq:U}
  \end{align}
\end{subequations} 
where at this stage we take~$\mathscr{F}\left(\TimeFunc\right)$ to be an arbitrary function of~$\TimeFunc$ whose only requirement is that~$U$ and~$V$ do not diverge as~$r \mapsto \Rh$. In these new coordinates, we compute 
\begin{align}\label{eq:dVdU}
	\mathrm{d}V \mathrm{d}U = \frac{\mathscr{F}^2}{4\Rh^2} \mathrm{d}t^2 - \left(\frac{\mathrm{d}\mathscr{F}}{\mathrm{d}\TimeFunc}\right)^2 \left(\frac{\mathrm{d}\TimeFunc}{\mathrm{d}r}\right)^2 \frac{\TimeFunc}{\SpaceFunc} \mathrm{d}\Bar{r}^2.
\end{align} 
Since we require the combination~$\mathrm{d}t^2 - \mathrm{d}\Bar{r}^2$ as in~\cref{eq:ds_rbar}, we find that~\cref{eq:dVdU} tells us that~$\mathscr{F}$ must satisfy
\begin{align}\label{eq:f_choice}
	\frac{\mathscr{F}^2}{4\Rh^2} = \left( \frac{\mathrm{d}\mathscr{F}}{\mathrm{d}\TimeFunc} \right)^2 \left(\frac{\mathrm{d}\TimeFunc}{\mathrm{d}r}\right)^2 \frac{\TimeFunc}{\SpaceFunc}.
\end{align} 
If~\cref{eq:f_choice} holds, then~\cref{eq:ds_rbar} can be rewritten as 
\begin{subequations}
  \begin{align}
	  \mathrm{d}s^2 &= \frac{4\Rh^2 \TimeFunc}{\mathscr{F}^2} \mathrm{d}V \mathrm{d}U\label{eq:ds_finalUV_up}\\
	  &= \SpaceFunc\left(\frac{\mathrm{d}\TimeFunc}{\mathrm{d}\mathscr{F}}\right)^2\left(\frac{\mathrm{d}r}{\mathrm{d}\TimeFunc}\right)^2 \mathrm{d}V \mathrm{d}U. \label{eq:ds_finalUV_down}
  \end{align}
\end{subequations} 
We demand that the metric to remain nonzero at~$r\mapsto\Rh$. This is equivalent to requiring 
\begin{align}\label{eq:condition}
	\lim_{\TimeFunc \mapsto 0}\left( \frac{\TimeFunc}{\mathscr{F}^2} \right) = \text{const.}
\end{align}
The constraint in~\cref{eq:condition} allows us to write~$\mathscr{F}(\TimeFunc)$ as 
\begin{align}\label{eq:f_expr}
	\mathscr{F}\equiv\sqrt{\TimeFunc}\sum_{n=0}^{\infty}\mathscr{F}_n\TimeFunc^{np_n},\quad \mathscr{F}_0\neq 0,\quad p_{n+1}>\frac{np_n}{n+1}.
\end{align} 
By rewriting~\cref{eq:f_choice}, we obtain 
\begin{align}\label{eq:condition2}
	\left[\frac{\mathrm{d}\left(\ln{\mathscr{F}}\right)}{\mathrm{d}\TimeFunc} \right]^{-1} = 2 \Rh \frac{\mathrm{d}\TimeFunc}{\mathrm{d}r} \sqrt{\frac{\TimeFunc}{\SpaceFunc}}.
\end{align} 
Meanwhile, taking~$r \mapsto \Rh$ (i.e.~$\TimeFunc(r) \mapsto 0$) in~\cref{eq:f_expr}, we already know
\begin{align}\label{dlnf_dT_rh}
	\lim_{\TimeFunc \mapsto 0}\left(\frac{\mathrm{d}(\ln{\mathscr{F}})}{\mathrm{d}\TimeFunc} \right) = \frac{1}{2\TimeFunc}.
\end{align} 
Substituting~\cref{dlnf_dT_rh} into~\cref{eq:condition2}, we obtain 
\begin{align}
	\lim_{r \mapsto \Rh}\left[\tensor{g}{^{ij}}\tensor{\partial}{_i}\tensor{g}{_{tt}}\tensor{\partial}{_j}\tensor{g}{_{tt}} \right] 
	&= \lim_{\TimeFunc \mapsto 0}\left[\frac{1}{\SpaceFunc}\left(\frac{\mathrm{d}\TimeFunc}{\mathrm{d}r}\right)^2\right] 
	\nonumber\\
	&= 4 \Rh \lim_{r \mapsto \Rh}\left[\TimeFunc\right] = 0,
\end{align} 
which confirms the condition in~\cref{eq:vollick}. While we have shown that~\cref{eq:vollick} and~$\TimeFunc(\Rh) = 0$ are equivalent, the Kruskal-like transformation is not always possible.

To determine the necessary conditions for a Kruskal-like transformation to be possible, we further investigate the properties of the radial metric function~$\SpaceFunc$ and the tortoise coordinate~$\Bar{r}$. First, if~$\mathrm{d}\TimeFunc/\mathrm{d}r$ remains finite and nonzero as~$r\mapsto\Rh$, then from~\cref{eq:ds_finalUV_down} and~\cref{eq:f_expr}, it follows that~$\SpaceFunc \sim 1/\TimeFunc$ in the vicinity of~$\Rh$. Next, for the tortoise coordinate, we rewrite~\cref{eq:f_choice} as 
\begin{align}\label{PreEquation}
	\frac{\mathscr{F}}{2\Rh} = \pm \frac{\mathrm{d}\mathscr{F}}{\mathrm{d}r} \frac{\mathrm{d} r}{\mathrm{d} \Bar{r}},
\end{align}
and~\cref{PreEquation} integrates to 
\begin{align}\label{eq:f_INo_rbar}
	\mathscr{F} = e^{\pm\Bar{r}/2\Rh}.
\end{align}
Now~\cref{eq:f_INo_rbar} implies that we require~$\Bar{r}\mapsto\pm\infty$ as~$r\mapsto\Rh$ if~$\TimeFunc(\Rh) = 0$. These two conditions determine whether a coordinate transformation into Kruskal-like form is possible. Thus,~\cref{eq:vollick} does not universally determine a null surface.

To illustrate this point, consider the metric
\begin{equation}\label{CounterExample}
	\mathrm{d}s^2 = r\ \mathrm{d}t^2 -r^3 \mathrm{d}r^2.
\end{equation}
In~\cref{CounterExample}, we see that~$r=\Rh=0$ is null since~$\TimeFunc(0) = 0$. A Kruskal-like transformation is not possible in this case, because the tortoise coordinate remains finite. Whilst a Kruskal-like transformation is ruled out, an EF-like transformation still allows us to identify the null quality of the surface. This example demonstrates that neither~\cref{eq:vollick} nor the condition~$\tensor{g}{^{rr}}=0$ apply in all cases.

\subsection{General static spacetime}\label{sec:GeneralS}
The analysis of~\cref{sec:2d} can be extended to a general (i.e. asymmetric) static spacetime for which the line element
\begin{align}\label{eq:general_metric}
  \mathrm{d}s^2\onshell \TimeFunc(r,\theta,\phi)\ \mathrm{d}t^2 + \tensor{g}{_{ij}}(r,\theta,\phi)\ \mathrm{d}\tensor{x}{^i} \mathrm{d}\tensor{x}{^j},
\end{align} 
replaces~\cref{eq:metric}. To apply the previous analysis, we note that it is always possible to perform a rotation into a coordinate system where the photon geodesic has no angular motion. In this new frame, the problem reduces to a two-dimensional one, allowing the earlier derivations to be applied. Since this argument applies to all points on the surface of interest, albeit with different transformations for different geodesics, we conclude that the necessary and sufficient condition for a surface to be null is~$\TimeFunc( \Rh)=0$ for general static spacetime. 

\subsection{Axisymmetric stationary spacetime}\label{sec:stationary}
It is also possible to extend the static analysis of~\cref{sec:2d,sec:GeneralS} to axisymmetric \emph{stationary} spacetimes. A key difference from the static case is the need to account for the angular motion of photon geodesics. We further extend~\cref{eq:general_metric} to the stationary line element
\begin{align}\label{eq:ds_stationary}
  \mathrm{d}s^2 &\onshell \AxiTimeFunc(r,\theta)\ \mathrm{d}t^2 + 2\MixTPhi(r,\theta)\ \mathrm{d}t \mathrm{d}\phi \nonumber\\
  &\quad - \SpaceFunc(r,\theta)\ \mathrm{d}r^2 - \Theta(r,\theta)\ \mathrm{d}\theta^2 - \Phi(r,\theta)\ \mathrm{d}\phi^2,
\end{align} 
where~$\SpaceFunc(r,\theta) \mapsto \infty$ for all $\theta$ as~$r\mapsto \Rh$. Without performing any coordinate transformations, one can rewrite~\cref{eq:ds_stationary} in the usual Boyer–-Lindquist form
\begin{align}\label{eq:BLform}
  \mathrm{d}s^2 &\onshell \left(\AxiTimeFunc(r,\theta) + \frac{\MixTPhi^2(r,\theta)}{\Phi(r,\theta)} \right) \mathrm{d}t^2 - \SpaceFunc(r,\theta)\ \mathrm{d}r^2 \nonumber\\ 
  &\quad - \Phi(r,\theta) \left( \mathrm{d}\phi - \frac{\MixTPhi(r,\theta)}{\Phi(r,\theta)} \mathrm{d}t \right)^2 - \Theta(r,\theta)\ \mathrm{d}\theta^2.
\end{align} 
The existence of an axial Killing vector implies that the angular momentum
\begin{align}
	L\equiv \MixTPhi \frac{\mathrm{d}t}{\mathrm{d}\lambda} - \Phi \frac{\mathrm{d}\phi}{\mathrm{d}\lambda},
\end{align} 
will be a conserved quantity along geodesics parameterised by some affine parameter~$\lambda$. We consider a radially infalling photon at spatial infinity, which we take to be asymptotically flat, so that~$L\onshell0$. This means that the photon follows a trajectory where 
\begin{align}\label{eq:dphi_dt}
  \frac{\mathrm{d}\phi}{\mathrm{d}t} \onshell \frac{\MixTPhi}{\Phi}.
\end{align} 
The condition in~\cref{eq:dphi_dt} once again simplifies the problem in~\cref{eq:BLform} to a two-dimensional system in which the effective line element function associated with~$t$ is 
\begin{align}\label{eq:eff_time}
  \TimeFunc &\equiv \AxiTimeFunc + \frac{\MixTPhi^2}{\Phi} \propto \det\left(\tensor{h}{_{ab}}\right),
\end{align} 
where~$\tensor{h}{_{ab}}$ is the induced metric on a hypersurface of constant~$r$ and~$\theta$. By applying the method of~\cref{sec:2d}, we conclude that a surface is null if and only if~$\TimeFunc\big(\Rh{},\theta\big) \onshell 0$ for all $\theta$, and this result agrees with the general condition~$\det\big(\tensor{h}{_{ab}}\big) = 0$. On the other hand, it should be noted that~\cref{eq:eff_time} demonstrates that the null surface does not generally coincide with the stationary surface, for which~$\AxiTimeFunc(r,\theta) \onshell 0$.

\subsection{Generalisations}\label{sec:generalisations}
The method proposed in~\cref{sec:2d,sec:GeneralS,sec:stationary} can be applied to other spacetimes. The key requirement is the ability to find first-order relations between the mixed metric components, such as that given in~\cref{eq:dphi_dt}. A simple example (though not a relevant solution to the Einstein field equations) is a spacetime where the metric is independent of~$\phi$ and~$\theta$. In this case, an additional Killing equation can be obtained, leading to a relation for~$\theta$ similar to~\cref{eq:dphi_dt}. Consequently, the null condition remains~$\TimeFunc \propto \det\big(\tensor{h}{_{ab}}\big) \onshell 0$, but~$\tensor{h}{_{ab}}$ is now the induced metric on a hypersurface of constant~$r$. For general spacetimes, it may be possible to derive the necessary first-order evolution relations using the Hamilton--Jacobi equation~\cite{PhysRev.174.1559}. Consequently, this approach may provide a pathway to further generalising our analysis. A detailed exploration of this extension is, however, beyond the scope of the present work.

\section{Examples}\label{sec:example}
Having developed the various null surface criteria across~\cref{sec:NS}, we now apply them to several examples of traversable wormholes.

\subsection{Ellis Drainhole}\label{sec:Ellis}
As a first example, we consider the Ellis drainhole~\cite{10.1063/1.1666161}, which is a static traversable wormhole. This solution can be further extended to model a bouncing cosmology~\cite{Ellis:2007dmt}. The line element is usually written as
\begin{align} \label{eq:ellis}
	\mathrm{d}s^2 = \left( 1-g\left(r\right)^2 \right) \mathrm{d} t^2 
	- \frac{ \mathrm{d} r^2 }{1-g\left(r\right)^2}
  - \rho\left(r\right)^2 \mathrm{d}\Omega^2,
\end{align}

where the functions~$g(r)$ and~$\rho(r)$ are determined by the field equations. Note that~\cref{eq:ellis} does not necessarily conform to the Schwarzschild-like coordinates in~\cref{eq:metric}, since the enclosing spheres do not have a proper area of~$4\pi r^2$. So long as~$\rho(r)$ remains finite, the condition worked out in~\cref{sec:2d} is insensitive to the scaling of~$r$, and so we conclude that a null surface exists if and only if~$g\left(\Rh\right)\onshell \pm 1$. This special case corresponds to the Schwarzschild solution. In general, for an Ellis drainhole,~$g(r)^2 < 1$, implying the absence of a horizon or null surface.

\begin{figure*}[ht!]
\includegraphics[width=\textwidth]{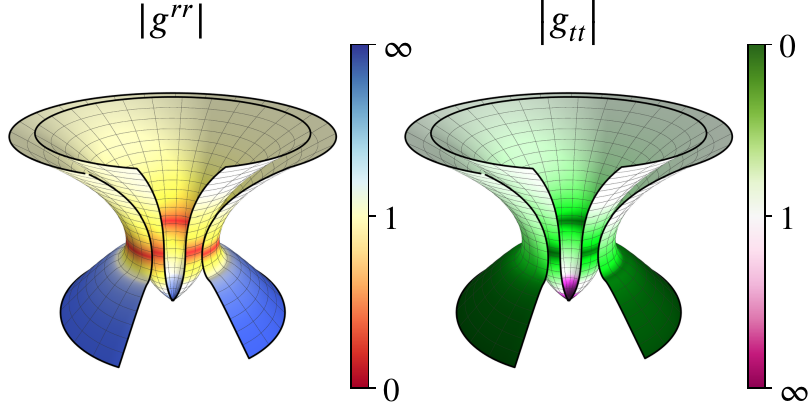}
	\caption{\label{fig:Wormhole} Different metric functions (left and right) are used to color cylindrical embeddings of an Eling--Jacobson wormhole with~$\Kb{}=1$ (outer surfaces) for comparison with a Schwarzschild black hole of the same asymptotic mass (inner surfaces). The upper halves show the asymptotically flat wormhole near side and black hole exterior, which would share the same geometry at infinity (here truncated). The wormhole throat lies at a dilated radial coordinate compared to the black hole horizon, and on this diagram the throat and horizon are aligned vertically. The condition~$\tensor{g}{^{rr}}=0$ is met both at the wormhole throat and black hole horizon. The condition~$\tensor{g}{_{tt}}=0$ is met only at the black hole horizon, and so the throat of the wormhole is not a null surface. The lower halves show the `flare-out' of the wormhole far side and the `pinch-off' of the black hole interior at the singularity. A curvature singularity would also occur at infinite radius on the far side of the wormhole (here truncated). This singularity would be null, as evidenced by the steadily decreasing value of~$\tensor{g}{_{tt}}$ in the visible region.}
\end{figure*}

\subsection{Einstein-\ae{}ther wormhole}\label{sec:EAE}

In this section, we consider the Eling--Jacobson wormhole. This spacetime was originally identified as a solution to the field equations of minimal Einstein-\ae{}ther (E\AE{}) theory with a Maxwell-like kinetic term~\cite{Clifton:2011jh,Jacobson:2000xp,Eling:2003rd,Jacobson:2004ts,Eling:2004dk,Foster:2005dk,Eling:2006ec,Konoplya:2006rv,Eling:2007xh,Jacobson:2007veq,Garfinkle:2007bk,Tamaki:2007kz,Barausse:2011pu,Berglund:2012fk,Berglund:2012bu,Gao:2013im,Barausse:2015frm,Campista:2018gfi,Bhattacharjee:2018nus,Oost:2018tcv,Lin:2018ken,Zhang:2019iim,Zhu:2019ura,Leon:2019jnu,Coley:2019tyx,Zhang:2020too,Adam:2021vsk,Chan:2022mxd} (see also the review~\cite{Monfort-Urkizu:2023jop}). Minimal E\AE{} theory is not considered viable due to instabilities, but it has recently been shown~\cite{Skordis:2024wlo,Yang:2025} that all the static spherically symmetric solutions to E\AE{} theory are automatically inherited by a more promising modified gravity model known as \ae{}ther-scalar-tensor theory (\AE{}ST)~\cite{Skordis:2020eui,Bataki:2023uuy,Skordis:2021mry,Verwayen:2023sds,Llinares:2023lky,Mistele:2023paq,Mistele:2023fwd,Bernardo:2022acn,Tian:2023gjt,Kashfi:2022dyb,Rosa:2023qun}, a relativistic completion of modified Newtonian dynamics (MOND) which is thought to be phenomenologically consistent with precision cosmology.\footnote{Note that the specific version of the \AE{}ST action for which this correspondence holds is one for which the MOND phenomenology is actually disabled. Moreover, the stability of such solutions against perturbations has not been worked out in the \AE{}ST context.} Unlike the case in~\cref{eq:ellis}, Schwarszschild-like coordinates as in~\cref{eq:metric} are commonly used for the Eling--Jacobson solution. In this case, however, the wormhole is asymmetric: the two possible branches of the time function $\TimeFunc{}$ are given, not in terms of~$r$, but in terms of~$\SpaceFunc$ by the formula
\begin{align}\label{eq:time_norm}
	\TimeFunc_{\pm}(\SpaceFunc) &=
	\left[ \left(\frac{\sqrt{2}-\sqrt{2-\Kb}}{\sqrt{2}+\sqrt{2-\Kb}}\right)^{\pm1} \right. \nonumber \\ 
	&\hspace{-10pt} \times \left .\left(\frac{\sqrt{2+\Kb(\SpaceFunc-1)}-\sqrt{2-\Kb}}{\sqrt{2+\Kb (\SpaceFunc-1)}+\sqrt{2-\Kb}}\right) \right]^{\pm\sqrt{\frac{2}{2-\Kb}}}.
\end{align}
In~\cref{eq:time_norm} the dimensionless constant~$0 < \Kb < 2$ is termed the \emph{\ae{}ther coupling}, and it is the only parameter (apart from the Planck mass or Newton--Cavendish constant) that appears in the E\AE{} theory action~\cite{Hsu:2024ftc}. The coupling~$\Kb$ is also a parameter in \AE{}ST, in which it takes the same numerical range, and in both theories this range is associated with the stability of vector perturbations. Accompanying the formulae in~\cref{eq:time_norm}, the two corresponding branches of the Schwarzschild-like radial coordinate~$r$ are also given as a function of~$\SpaceFunc$ by
\begin{align}\label{RadiusInTermsOfSpaceFunc}
r_{\pm}\left(\SpaceFunc\right)\onshell & \frac{\Rh\sqrt{2\Kb\SpaceFunc}}{\sqrt{4+2\Kb\left(\SpaceFunc-1\right)}\pm2}
	\nonumber\\
	&\hspace{-10pt} \times\left[\frac{\sqrt{2+\Kb\left(\SpaceFunc-1\right)}+\sqrt{2-\Kb}}{\sqrt{2+\Kb\left(\SpaceFunc-1\right)}-\sqrt{2-\Kb}}\right]^{\pm\frac{1}{\sqrt{4-2\Kb}}}.
\end{align}
In~\cref{RadiusInTermsOfSpaceFunc} the branches~$r_{+}$ and~$r_{-}$ respectively cover the \emph{far} and \emph{near} sides of a wormhole whose throat lies at~$r_+=r_-=\Rh$. The near side has the property of asymptotic flatness in the sense that~$r_-\big(\SpaceFunc\big)\mapsto \infty~$ as~$\SpaceFunc\mapsto 1$, and~$\TimeFunc_{-}(1)=1$. Moreover, the asymptotic behavior of the near-side spacetime is identical to the Newtonian limit of a Schwarzschild black hole, whose Schwarzschild radius (or Newtonian `weight') is related to~$\Rh$ by some~$\Kb$-dependent parameterisation. This parameterisation is such, that as~$\Kb\mapsto 0$ the whole near-side solution approaches Schwarzschild spacetime. At all finite~$\Kb$, however, the nonlinear core of the spacetime actually has quite a different structure, and in particular it always lacks a horizon. To understand this property, notice how, as~$\SpaceFunc\mapsto \infty$, it follows from~\cref{RadiusInTermsOfSpaceFunc} that~$r_\pm\big(\SpaceFunc\big)\mapsto\Rh$ from either side. As discussed in~\cref{sec:intro}, one might erroneously conclude from this observation that~$r_+=r_-=\Rh$ is a horizon owing to the property of being null. However, by inspecting~\cref{eq:time_norm}, we can see that~$\TimeFunc_{\pm}\big(\SpaceFunc\big)$ remains finite in this limit, so the throat is not actually a null surface. 

By contrast with the behaviour at the throat,~\cref{RadiusInTermsOfSpaceFunc} reveals that the far-side spatial infinity~$r_+\big(\SpaceFunc{}\big)\mapsto \infty~$ corresponds to the limit~$\SpaceFunc\mapsto 0$. Turning to~\cref{eq:time_norm}, it can be found that~$\TimeFunc_+\big(0\big)=0$. It is confirmed in~\cite{Yang:2025} that this surface is a null singularity, i.e. a null surface on which various curvature scalars diverge.

We illustrate these results in~\cref{fig:Wormhole}, where the properties of the Eling--Jacobson wormhole are compared with those of a Schwarzschild black hole. Focussing on spatial foliations, this diagram embeds 2D equatorial~$\theta=\pi/2$ slices (for which $\mathrm{d}\Omega=\mathrm{d}\phi$) spanned by radial~$r$ and azimuthal~$\phi$, in a 3D cylindrical coordinate system spanned by axial~$z$, radial~$r$ and azimuthal~$\phi$. Working from~\cref{eq:metric} we identify
\begin{align}\label{EmbeddingLineElement}
	\SpaceFunc\ \mathrm{d}r^2 + r^2 \mathrm{d}\phi^2 = \text{sgn}\left[\SpaceFunc-1\right]\mathrm{d}z^2 + \mathrm{d}r^2 + r^2 \mathrm{d}\upphi^2,
\end{align} 
where the signature of the embedding space on the RHS changes abruptly according to the value of~$\SpaceFunc$, such that~\cref{EmbeddingLineElement} yields
\begin{equation}\label{EmbeddingCriterion}
	\frac{\mathrm{d}z}{\mathrm{d}r} = \sqrt{\left|\SpaceFunc-1\right|}.
\end{equation}
This change in signature extends~\cref{EmbeddingCriterion} formally to cover regions where~$\SpaceFunc<1$, such as happens for the whole of the interior of the Schwarzschild black hole. This condition is also encountered some radial distance into the far side of the Eling--Jacobson wormhole, and it persists as $\SpaceFunc{}\mapsto 0$ all the way to the null singularity at~$r_+\big(\SpaceFunc{}\big)\mapsto \infty~$.

\subsection{Rotating traversable wormholes}
Rotating Teo wormholes\footnote{See also~\cite{PhysRevLett.132.121501,Al-Balushi:2019aa} for related examples.}~\cite{PhysRevD.58.024014} are constructed following the procedure used for static Morris--Thorne wormholes~\cite{10.1119/1.15620}. Teo wormholes have been commonly used to model axisymmetric traversable wormholes. Using Boyer--Lindquist coordinates as in~\cref{eq:BLform}, the Teo line element is given by 
\begin{align}
  \mathrm{d}s^2 &\onshell N^2\left(r,\theta\right) \mathrm{d}t^2 
  - \left[1-\frac{b\left(r,\theta\right)}{r}\right]^{-1} \mathrm{d}r^2 \nonumber\\ 
	&\quad - r^2 K^2 \left(r,\theta\right) \left[ \sin^2(\theta)\left(\mathrm{d} \phi - \omega(r,\theta) \mathrm{d}t \right)^2 + \mathrm{d}\theta^2 \right],
\end{align}
where the \emph{shape function}~$b(r,\theta)$ satisfies~$b \leq r$ and is independent of~$\theta$ at the throat, where~$\Rh\equiv b\big(\Rh,\theta\big)\equiv b\big(\Rh\big) > 0$. The function~$K(r,\theta)$ is a positive, nondecreasing function of~$r$ that determines the proper radial distance, while~$\omega(r,\theta)$ encodes the angular velocity of the wormhole. For the wormhole to be traversable, the \emph{redshift function}~$N(r,\theta)$ must remain finite and nonzero throughout the spacetime, ensuring the absence of event horizons or curvature singularities. This condition thereby confirms that the throat is not a null surface. 

\section{Conclusion}\label{sec:conclusion}
In this letter, we have presented a method to identify null hypersurfaces via coordinate transformations. By employing Eddington--Finkelstein-like coordinates, we recover the criterion that for a surface to be null in a spherically symmetric static spacetime one requires the timelike metric component to vanish, i.e.,~$\tensor{g}{_{tt}}=0$ on the surface~\cite{10.1063/1.1664717,Vollick:2015aa}. We further extended this analysis using a Kruskal-like coordinate transformation, where~\cref{eq:vollick} was recovered. However, we showed that such Kruskal-like coordinates are not always attainable. Specifically, if~$\tensor{\partial}{_r}\tensor{g}{_{tt}} \neq 0$, the ratio~$\tensor{g}{_{tt}}\tensor{g}{^{rr}}$ must remain finite and nonzero in the vicinity of the surface of interest. We then generalised our analysis to asymmetric static spacetimes, for which we demonstrated that the problem can always be locally reduced to a two-dimensional form through an appropriate coordinate transformation. Consequently, for any static spacetime, the nullity condition remains~$\tensor{g}{_{tt}}=0$. 

For axisymmetric stationary spacetimes, the non-trivial angular motion of photons near the horizon necessitates a more refined approach. By expressing the metric in Boyer--Lindquist coordinates, we isolated the angular motion and applied the two-dimensional analysis. We found that the nullity condition is given by~$\tensor{g}{_{tt}}=0$ for a reduced line element, as defined in~\cref{eq:eff_time}. This result is consistent with the more general condition that the determinant of the induced metric on the hypersurface must vanish, i.e.,~$\mathrm{det}\big(\tensor{h}{_{ab}}\big)=0$.

Our analysis provides a systematic method for identifying null surfaces in both static and axisymmetric stationary spacetimes, which may possibly be extended to more general cases, and offers insights into the underlying connection between such surfaces and null geodesics. 

\begin{acknowledgements}
We are grateful for useful discussions with Amel Durakovic, Justin Feng, Tobias Mistele, Constantinos Skordis, David Vokrouhlicky and Lirui Yang.

Y-HH is supported by the doctoral scholarship from the Taiwan Ministry of Education. WB is grateful for the support of Marie Skłodowska-Curie Actions and the Institute of Physics of the Czech Academy of Sciences.

    Co-funded by the European Union. Views and opinions expressed are however those of the author(s) only and do not necessarily reflect those of the European Union or European Research Executive Agency. Neither the European Union nor the granting authority can be held responsible for them.
\end{acknowledgements}

\bibliographystyle{apsrev4-1}
\bibliography{NotINSPIRE}

\end{document}